\documentclass[%
superscriptaddress,
amsmath,amssymb,
aps,
prb,
twocolumn,
]{revtex4-1}

\usepackage{graphicx}
\usepackage{ulem}
\usepackage[]{latexsym}
\usepackage{bm}
\usepackage{color}

\newcommand{\mb}[1]{\mbox{\boldmath $#1$}}

\newcommand{\eas}[0]{\begin{eqnarray*}}
\newcommand{\eae}[0]{\end{eqnarray*}}
\newcommand{\les}[0]{\begin{equation}}
\newcommand{\lee}[0]{\end{equation}}
\newcommand{\leas}[0]{\begin{eqnarray}}
\newcommand{\leae}[0]{\end{eqnarray}}
\newcommand{\mchss}[4]
{
\left\{
\begin{array}{cc}
#1 & #2   \\
#3 & #4
\end{array}
\right.
}

\newcommand{\mmat}[4]
{
\left(
\begin{array}{cc}
#1 & #2 \\
#3 & #4 
\end{array}
\right)
}
\newcommand{\lmat}[4]
{
\left|
\begin{array}{cc}
#1 & #2 \\
#3 & #4 
\end{array}
\right|
}

\newcommand{\mvec}[2]
{
\left(
\begin{array}{c}
#1  \\
#2  
\end{array}
\right)
}

\newcommand{\mynote}[1]{}
\newcommand{\mcr}[1]{{#1}}

\begin{document}



\title{
%
  {
    Survival of sharp $n=0$ Landau levels 
  in massive tilted Dirac fermions:\\
  Role of the generalized chiral operator
}
}


\author{Yasuhiro Hatsugai}
\affiliation{Institute of Physics, University of Tsukuba, Tsukuba, 305-8571 Japan}

\author{Tohru Kawarabayashi}
\affiliation{Department of Physics, Toho University,
Funabashi, 274-8510 Japan}

\author{Hideo Aoki}
\affiliation{Department of Physics, University of Tokyo, Hongo, 
Tokyo 113-0033 Japan} 
\affiliation{High Energy Accelerator Research Organization (KEK), Tsukuba,
Ibaraki 305-0801 Japan}



\date{\today}

\begin{abstract}
Anomalously sharp  (delta-function-like) $n=0$ Landau level in the presence of disorder is usually considered to be a manifestation of the 
massless Dirac fermions in magnetic fields.  
This property persists even when the Dirac cone is tilted, 
which has been shown by Kawarabayashi et {
  al. [Phys. Rev. B 
{\bf 83}, 153414 (2011)]} to be a 
consequence of a ``generalized chiral symmetry".  
Here we pose a question whether this property will be washed out when the 
tilted Dirac fermion becomes massive.  
Surprisingly, the levels persist to be delta-function-like, 
although the mass term that splits 
$n=0$ Landau levels may seem to degrade the anomalous 
sharpness.  
This has been shown both numerically for a tight-binding 
model, and analytically in terms of the Aharonov-Casher argument 
extended to the massive tilted Dirac \mcr{fermions}.  
A key observation is that, 
while the generalized chiral symmetry is broken by the mass term, the $n=0$ Landau level remains to accommodate eigenstates of the generalized chiral operator, resulting in the robustness against chiral-symmetric disorders.  
Mathematically, the conventional and generalized chiral operators 
are related with each other via a non-unitary transformation, 
with which the split, nonzero-energy $n=0$ wave functions of the massive system are just 
gauge-transformed zero-mode wave functions of the massless 
system.  
A message is that the chiral symmetry, rather than a simpler notion 
of the sublattice symmetry, {
  is essential for} 
 the robustness of the $n=0$ Landau level.
\end{abstract}


\maketitle

\renewcommand{\widetilde}[1]{^t\!#1}

\section{Introduction}
\label{sec:intro}
After the experimental discovery of graphene,\cite{Novo05,Novo05} 
fascinations with the 
massless Dirac fermions have become one of the central interests 
in condensed-matter physics. {
  \cite{Aoki14}}
Physics of zero-gap semiconductors has actually a 
long history of studies, started by
 a theoretical work by Wallace and 
is now described in condensed-matter 
textbooks.\cite{Harrison89} 
There are various spin-offs, among which is the topological insulator 
with the quantized spin Hall effect, 
where the topological property of Dirac fermions plays a fundamental 
role.\cite{Kane05,Moore10,Hasan10,RMPZhang11}  
In the context of zero-gap semiconductors, 
the first topological insulator, HgTe-CdTe, was realized 
by chaging the mass to be negative.\cite{Harrison89,Bernevig06} 
Quantum phase transitions of fermions 
associated with gap closing and opening can be described by 
a Dirac fermion in terms of reversing the sign of the mass.{
  \cite{Hats96}} We have then realizations of 
diverse quantum phases, such as chiral spin states, 
flux phases, and nodal fermions. 
Another important class of the Dirac fermions is an organic 
material,  
$\alpha$-(BEDT-TTF)$_2$I$_3$\cite{Kata06,Koba07,kajita14},
where the Dirac cone dispersion is substantially tilted.
In a broader context, anisotropic superconductors with d-wave symmetry 
has Dirac cones in the dispersion for the Bogoliubov quasi-particle, 
which serve as another Dirac fermions
in two dimensions.{
  \cite{Lee93,PhysRevB.48.4204}
}

{
  While the massless Dirac cone in graphene is  
related to the honeycomb lattice 
structure, the gap closing itself can be analyzed more generally
in terms of the level crossing in quantum mechanics. 
According to 
the von-Neumann Wigner theorem, a degeneracy point has generically 
co-dimension three.\cite{Berry84,Herring37,Hatsugai10S}  
This indicates that the existence of massless Dirac cones in three spatial 
dimensions is rather natural. Conversely, 
in two dimensions a Dirac cone is an accident unless 
some symmetry exists.
The chiral symmetry{
  \cite{Hatsbook14}} 
is often evoked for graphene as represented 
by the honeycomb lattice, for which the symmetry 
is usually regarded as nothing but the sublattice symmetry 
against {
  sign change of the wave function on one of the
sublattices in a bipartite 
lattice structure.}  
Hence it is a usual practise to attribute the reason why graphene 
realizes the massless Dirac fermions to the honeycomb structure.  
In two-dimensional systems with a chiral symmetry, one can 
also prove 
the fermion doubling theorem as a two-dimensional analogue of
the Nielsen-Ninomiya theorem conceived for four dimensions,\cite{NNT,HFAstability,HatsuSSC,Hatsugai10S} 
which dictates the number of Dirac cones to be even.
{
  It also brings a supersymmetric (SUSY) structure in the one-particle Hamiltonian.\cite{Witten82,Ezawa07,EPLSUSY09}}
In the case of graphene 
this is why we have two Dirac cones at valleys K and K'. 
Thus in the physics of graphene the chiral
symmetry is important.\cite{Hatsugai06-Gra,HFAstability,HatsuSSC,Kawarabayashi09,Hatsbook14}
We can even use the chiral 
symmetry to discuss topological nature of the system.{
  \cite{Hatsbook14,Hatsugai06-Gra}} For instance, 
in a d-wave superconductor, the chiral symmetry 
translates into the time-reversal symmetry
in the Bogoliubov Hamiltonian,\cite{Ryu02,HatsuSSC} which protects
the existence of nodes in the gap.

Now, in two dimensions the Dirac cone is {
  {\it in general}} tilted, 
as in the case of the organic material, 
where the conventional chiral symmetry is broken.\cite{Kawa11}  
One may then wonder if the existence of a Dirac cone itself 
suffices for the topological properties 
even when the chiral symmetry is apparently absent.  
The present authors have revealed that the notion 
of the chiral symmetry can actually 
be extended to accommodate 
the tilted cones,\cite{Kawa11}
where the tilted cones has a symmetry 
against the ``generalized chiral operator", 
and demonstrated some of 
its consequences both analytically and numerically. 
Most importantly, if we look at $n=0$ Landau level 
(right at the Dirac point) in magnetic fields, 
its density of states remains 
delta-function-like even in the presence of 
disorder, while one might assume that this property would be 
specific to vertical Dirac cones.  

In the present paper 
we pose a new question whether the anomalous property of the 
$n=0$ Landau level will be washed out when the tilted Dirac fermion 
becomes massive.  
While this question may seem too detailed, it is actually 
not so, since from this we can clarify an important 
question: are the existence of zero-modes and the chiral symmetry one and the same?  
While for a vertical cone they are obviously the same, 
in a massive case the $n=0$ Landau level splits into two with 
nonzero energies, so that one might 
imagine that the two properties should differ from 
each other in this case.  
Surprisingly, we shall find that the levels, now split, 
do remain delta-function-like.  
This has been shown analytically 
in terms of the Aharonov-Casher argument, which is 
known to 
construct wave functions in the zero-mode Landau level 
in vertical cones, and is here 
extended to the massive tilted Dirac fermions.  
A key observation is that, 
while the generalized chiral {\it symmetry} is broken by the mass term, the $n=0$ Landau level remains to accommodate eigenstates of the generalized chiral {\it operator}, ensuring the robustness against chiral-symmetric disorders.  
Mathematically, the conventional and 
generalized chiral operators 
are found to be related with each other via a non-unitary transformation, 
with which we can identify 
the split, nonzero-energy wave functions of the massive system
as a 
{\it gauge-transformed zero-mode wave functions} of the massless ones.  
The anomalously sharp Landau level 
is confirmed from a numerical result for a model tight-binding system 
for disorders that respect the generalized-chiral symmetry, in 
sharp contrast with the disorders that do not. 

We can visualize the point as follows.  
While the conventional chiral symmetry dictates that 
each wave function in the $n=0$ Landau level has nonzero 
amplitudes only on A sublattice (or B sublattice), the wave function 
for tilted Dirac fermions is not an eigenstate of the sublattice 
symmetry, so that it 
has amplitudes on both of the two sublattices 
(or two components of the spinor).  
This may seem to suggest that the sharpness of the $n=0$ Landau level is 
degraded for tilted Dirac fermions when we make the 
fermion massive by introducing a staggered potential over A and B 
sublattices. 
If this is the case, the tilted cone should differ 
from the vertical cone, and the sharpness of the $n=0$ Landau level 
will be affected by the staggered potential.  
The present result shows that this is not the case.  
Thus a message of the present work 
is that (i) the (generalized) chiral symmetry rather than a simpler notion 
of the sublattice symmetry is essential
for the robustness of the $n=0$ Landau level,
which is why (ii) the chiral
{
  operator}
{
  plays a crutial role}
even {
  in the } massive case. 

{
  Since the presence of Dirac cones  is accidental in 2D 
systems unless there is some symmetry protection, it is natural to expect an energy gap in  Dirac fermion systems. 
  Hence the massive Dirac fermion with tilting  in 2D
  is a generic and common problem.
{
  In fact, 
  the extensive studies are now going on for massive Dirac-fermion materials such as molybdenum disulfide compounds\cite{mos}, as well as 
  several organic materials with 
 substantially tilted Dirac cones. 
 The insulating phase in such organic materials can be a candidate of the massive and tilted Dirac fermions.\cite{kajita14,doi:10.1143/JPSJ.74.2897,doi:10.7566/JPSJ.83.094701}
  In a completely different area, the massive and tilted Dirac fermions may be realized in cold atoms in optical lattices, where the Dirac cones are often tilted and the parameters are more controllable than in solid-state
  materials\cite{PhysRevB.82.241403,movedDirac,PhysRevLett.103.035301,PhysRevA.84.023622}.
}
} 
{
}


In this paper we start in section II with a numerical result for a lattice model 
that has tilted Dirac cones, where we find  
the anomalous sharpness of the $n=0$ Landau levels is surprisingly 
unaffected by the introduction of the mass term for the case of the 
spatially smooth (long-range) disorder, as far as the 
disorder respects the 
chiral symmetry (as is the case with random magnetic fields introduced 
there).  
We further find numerically that, 
for spatially uncorrelated (short-range) disorder, 
the anomalous sharpness of the $n=0$ Landau levels is unexpectedly recovered as the staggered potential 
is increased.  This is just the {\it opposite} to the case of 
massless cones, where the sharpness of the 
$n=0$ Landau level is degraded for spatially uncorrelated 
disorder.\cite{Kawa11} 
The recovery of the sharpness of the $n=0$ Landau level for the uncorrelated disorder has 
also been reported for the shifted Dirac cones.\cite{Kawa13} 
In the present case, the energies of the two $n=0$ Landau levels associated with the two valleys are split by the mass term (i.e. the staggered 
potential), although the Dirac cones themselves are not shifted in energy.  
The present recovery of the sharpness thus indicates that 
the disorder-induced mixing between the split 
$n=0$ Landau levels is significantly 
suppressed by increasing the energy splitting introduced by the mass term for a chiral-symmetry preserving disorder.  
This is in sharp contrast with a potential disorder as 
we also demonstrate numerically.


We then present in section III the main, analytic part, which provides a 
solution for the puzzling numerical result.  Namely, 
in order to understand the origin of the anomalous sharpness of the $n=0$ Landau level for a massive 
Dirac fermions, we develop a 
general effective theory, first for the massless case  
and for the massive case in section IV, and find a simple algebraic relationship 
(which turns out to be non-unitary) between 
the generalized chiral operator and the conventional 
chiral operator. This enables us to discuss quantitatively the effect of the staggered potential on 
the anomalous sharpness of the $n=0$ Landau levels for tilted Dirac fermions with
a disorder that respects the generalized chiral symmetry.
We then show that the $n=0$ Landau levels of the massive Dirac fermions 
are still the eigenstate of the generalized chiral operator,
where the robustness for the random gauge field
is again 
shown analytically by the argument by Aharonov and Casher.
{
  Although  an a priori introduction of the  generalized chiral operator is given in
  the ref.\cite{Kawa13}, here we provide a transparent and logical derivation using a 4-dimensional
notation. This enables us to consistently describe
the massless and massive Dirac fermions with tilting.
Since the massive and massless Dirac cones occur  as semimetals and 
semiconductors in 2D, 
the compact 4-dimensional notation given in the paper
can be useful for
understanding of the Dirac fermion related physics.
}
{
  Section V is devoted to summary.}

\section{Numerical result for massive tilted Dirac fermions}

To examine the $n=0$ Landau level for massive and tilted Dirac fermions, 
let us first perform a 
numerical analysis based on the tight-binding lattice model \cite{Kawa11}
on a two-dimensional square lattice,
{
  with a tight-binding 
Hamiltonian given by
\begin{eqnarray*}
 H_{TB} &=& \sum_{\bm{r}} \bigg[-t c_{\bm{r}+\hat{\bm{y}}}^\dagger c_{\bm{r}}+ (-1)^{x+y}t c_{\bm{r}+\hat{\bm{x}}}^\dagger c_{\bm{r}}+{\rm h.c.}\\
    & & + t' \left(c_{\bm{r}+\hat{\bm{x}}+\hat{\bm{y}}}^\dagger c_{\bm{r}} + c_{\bm{r}+\hat{\bm{x}}-\hat{\bm{y}}}^\dagger c_{\bm{r}} \right)+ {\rm h.c.}\bigg].
\end{eqnarray*}
Here the lattice positions are denoted by $\bm{r} = x \hat{\bm{x}} + y\hat{\bm{y}}$ with $\hat{\bm{x}}(\hat{\bm{y}})$ being the unit vector in the 
$x(y)$-direction and the length in units of the lattice constant of the square lattice, $t$ the nearest-neighbor hopping, and $t'$ the next nearest-neighbor hopping.  The model is similar to the $\pi$ flux model\cite{Hatsugai90}} that has a {
half flux per plaquette
  ($\pi$ flux)}.  
The factor $(-1)^{x+y}$ in the nearest-neighbor hopping 
is the Peierls phase for the half flux quantum.  When $t ^\prime =0$, the 
dispersion of the model
  has two massless Dirac cones at $E=0$ around the k-points 
$\bm{k}_0= (0,\pm \pi/2)$, while the Dirac cones become tilted 
when $t'\neq 0$.\cite{Kawa11}

{
The effective low-energy Hamiltonian $H$ around the Dirac cones  can then be expressed in terms of the Pauli matrices $\widetilde{\bm{\sigma }}=({\sigma _1},{\sigma _2},{\sigma _3}) 
\equiv ({\sigma _x},{\sigma _y},{\sigma _z}) 
$ as
$$ 
H =
 ( X^0\sigma _0  +\bm{X}\cdot  \bm{\sigma } )\delta k_x
+
 (Y^0 \sigma _0   +\bm{Y}\cdot \bm{\sigma } ) \delta k_y,
$$ 
where $\delta \bm{k}=\bm{k}-\bm{k}_0  $ is the deviation of
the momentum from the Dirac point $\bm{k}_0 $ and  
$\sigma _0$ a two-dimensional unit matrix.  
In the case of the above tight-binding Hamiltonian $H_{TB}$ 
we have $\widetilde{\bm{X}} = (0,2t,0)$ and $\widetilde{\bm{Y}}=(\mp 2t,0,0)$, and $(X^0,Y^0)=(0,\pm 4t')$ that makes the Dirac cone indeed tilted.  

Now we make the fermions massive by introducing a mass 
term.  
This can be readily done by introducing a staggered potential, and the massive Hamiltonian $H_{TB}(m)$ reads 
\begin{equation}
 H_{TB}(m) = H_{TB} + m{c^2}\sum_{\bm{r}} (-1)^{x+y}c_{\bm{r}}^\dagger c_{\bm{r}},
\label{lattice_model}
\end{equation}
where A (B) sublattice site-energies are elevated (lowered). 
The effective low-energy Hamiltonian becomes 
$$
 H(m) = H + mc^2 \sigma_z, 
$$
where the term $mc^2\sigma_z$ makes the Dirac fermions massive with a gap 
at the Dirac point. 
In the case of usual vertical Dirac fermions, the mass term can be expressed in terms of the conventional chiral operator $\Gamma \propto \sigma_z$. The chiral operator is generally defined 
as an operator that anti-commutes with the Hamiltonian $H$, which, for 
the
{
  vertical } 
Dirac cone with $X^0=Y^0 =0$, is given by $\Gamma=  \bm{\hat{n}}_0\cdot \bm{\sigma}$ with $\bm{\hat{n}}_0 \equiv \bm{X}\times \bm{Y}/|\bm{X}\times \bm{Y}|$.\cite{hatsugai2011,Kawa11}
For the present model, the conventional chiral operator is simply $\Gamma = \pm \sigma_z$, where 
the plus (minus) sign applies to the valley around $\bm{k}_0 = (0,\pi/2)$ $((0,-\pi/2))$. 
The Hamiltonian $H(m)$ is then expressed with $\Gamma$ as  
\begin{eqnarray}
  H(m) = &H& + mc^2 \Gamma \quad {
    {\rm for}\ \bm{k} = (0,\pi/2)}\nonumber\\
  &H& - mc^2 \Gamma \quad {
    {\rm for}\ \bm{k} = (0,-\pi/2)}.\label{eq:exValley}
\end{eqnarray}

We then apply an external magnetic field to 
carry out exact numerical diagonalization for a finite system in the presence of disorder.  
{
  The magnetic field is taken into account by the Peierls phase, $t(t') \to t(t')e^{2\pi i\theta}$, 
where the summation of the $\theta$ 
 along a loop is given by the enclosed magnetic flux, $\phi$, in units of the flux quantum $h/e$.}  
The disorder is introduced here as a random component, 
$\delta \phi(\bm{r})$, in the magnetic flux 
$\phi(\bm{r}) = \phi + \delta \phi(\bm{r})$
piercing each square plaquette, where $\phi$ denotes the uniform component.  
The random component $\delta \phi(\bm{r})$ is assumed to obey a Gaussian distribution with a variance $\sigma$ and a 
spatial correlation,
\[
\langle \delta \phi(\bm{r})\delta \phi(\bm{r}') \rangle = \sigma^2 \exp(-|\bm{r}-\bm{r}'|^2/4 d^2), 
\]
where  $d$ is the correlation length.  We have chosen a randomness in the 
magnetic field, since a disorder in gauge degrees of freedom (such as 
the random magnetic field) respects the generalized chiral symmetry,\cite{Kawa11}.

In Fig. \ref{fig:1}, 
we show the density of states of the system  with tilted  Dirac cones 
in a finite magnetic field 
($\phi/(h/e)=0.01$) for the case of the spatially correlated 
disorder ($d=1.5$ in units of the lattice constant). 
For comparison, we also display the result for the case of 
vertical Dirac cones.  We can immediately notice 
that the introduction of the mass term ($mc^2\sigma_z$) 
does not affect the anomalous 
sharpness of the split $n=0$ Landau levels even 
for the tilted cones as in the vertical cones.  Since the other Landau levels 
(e.g. $n=\pm 1$) are broadened, the $n=0$ levels do stand out. 

A further surprise occurs when we 
examine the robustness of the split $n=0$ Landau levels against the spatially uncorrelated disorder ($d/a=0$).  
For the massless ($m=0$) case, uncorrelated disorder degrades the 
sharp $n=0$ Landau levels due to the inter-valley scattering.\cite{Kawarabayashi09,Kawa12}  However, we can see in 
Fig. \ref{fig:2} that the anomalous sharpness is actually {\it recovered as the mass is made heavier} 
with the level splitting becoming wider. 
In the massive case, each $n=0$ Landau level is associated with one of the two Dirac cones. 
(See Fig.5 below.)
The present result 
indicates that the mixing between 
the Dirac cones is effectively suppressed when the $n=0$ Landau levels are split by the 
staggered potential.   
This reminds us of our previous work, where 
we have introduced a model in which 
the two Dirac cones remain massless but shifted in 
energy with a complex hopping.  There, 
the robustness is recovered even for short-range disorder.\cite{Kawa13} The present result indicates that a 
similar suppression of the mixing is at work, where the 
energy offset comes not from the shifted cones but 
from a mass gap.

{
  We can show that the situation becomes completely 
different for a spatially uncorrelated potential disorder 
which does not respect the generalized chiral symmetry.  
  For the present lattice model, a potential disorder 
can be represented by random site-energies
as $\sum_{\bm r} \varepsilon_{\bm r} c^\dagger_{\bm r} c_{\bm r}$ in 
place of the random component of the magnetic field. 
We then find, as clearly shown in Fig.3, that the recovery of the anomalous  sharpness of the $n=0$ Landau level is 
completely absent for the case of the potential disorder, even though the mixing between two valleys is suppressed by the mass term.  
This suggests that, although the mass term (staggered potential) naively breaks the generalized chiral symmetry of the system, 
whether the disorder respects the generalized chiral symmetry persists 
to be crucial for the anomalous sharpness of the 
$n=0$ Landau levels of the massive tilted Dirac fermions.
}

\begin{figure}
\includegraphics[width=8cm]{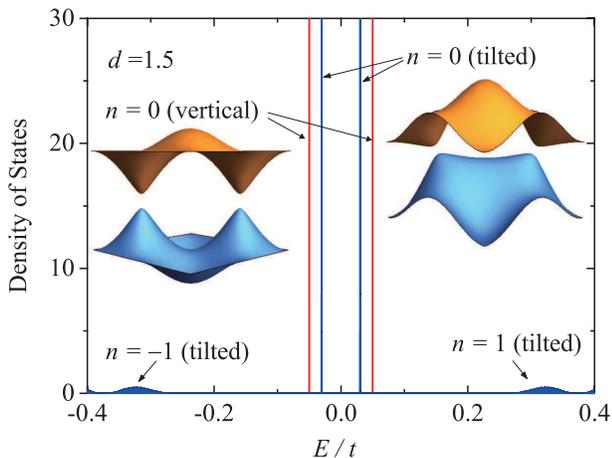}
\caption{(Color Online) Density of states for the lattice model having tilted Dirac cones in the presence of a 
mass term (staggered potential) in a magnetic field 
($\phi/(h/e)=0.01$) with a 
spatially correlated disorder ($d = 1.5$). 
Landau levels both for tilted ($t'/t=0.4$, blue) and for vertical ($t'/t=0$, red) cones are shown. 
The amplitude of disorder is taken here to be 
$\sigma/(h/e)= 0.0029$, the mass $m c^2/t = 0.05$, and the system-size 30 $\times$ 30 with an average over 5000 samples. 
 }
\label{fig:1}
\end{figure}

\begin{figure}
\includegraphics[width=9cm]{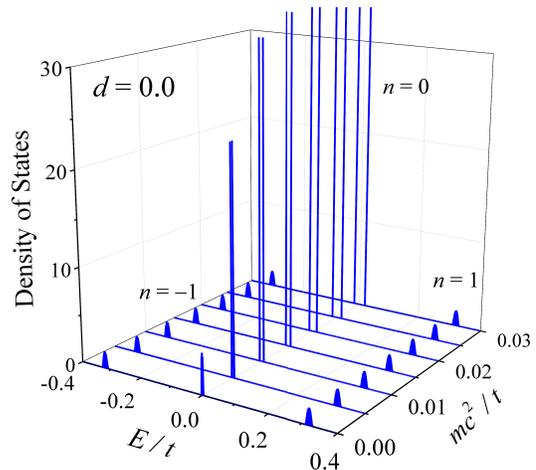}
\caption{For the spatially uncorrelated disorder ($d = 0$), 
the density of states 
for the lattice model having tilted Dirac cones is plotted 
against the mass (staggered potential) for the same 
magnetic field and the amplitude of disorder as in Fig.1. 
 }
\label{fig:2}
\end{figure}

\begin{figure}
\includegraphics[width=9cm]{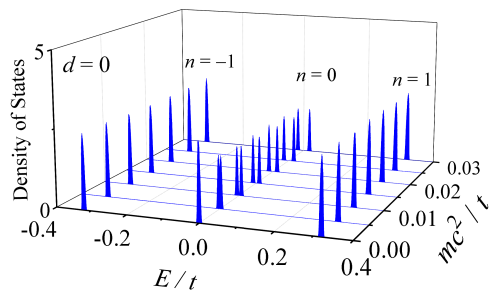}
\caption{ 
  The density of states for uncorrelated potential disorder, instead of 
the random magnetic field,  is plotted 
against the mass (staggered potential) for the same uniform
magnetic field as in Figs.1 and 2. The random potential is assumed to obey a 
Gaussian distribution with zero mean and a variance of $0.02 t$.
 }
\label{fig:2.5}
\end{figure}

\section{Tilted massless Dirac fermions}
\label{sec:tilted}

\subsection{A general formulation}

To understand the robustness of  the zero modes for massive and tilted Dirac fermions\cite{Goerb08,Mori09,Mori10},
let us first summarize a general effective theory for a massless and tilted Dirac fermions
from the viewpoint of the generalized chiral symmetry. 
For this purpose we  introduce a compact 
four-dimensional notations to make the discussion transparent.
{
  As a generic band structure of semiconductors, 
let us  consider a 
two-band Hamiltonian, $H_g$, in a $2\times2$ matrix form 
focusing on the valence and conduction bands.
Since the Hamitonian is hermitian, it is expanded by $\sigma_0,\sigma_1,\sigma_2$ and $\sigma_3$ as}
\begin{eqnarray*} 
{H}_g(\bm{k} ) &=&
\sigma _0 R_{0}(\bm{k} )
+
 \bm{\sigma } 
\cdot
  \bm{R} (\bm{k} ),
\end{eqnarray*} 
where
$\widetilde {\bm{R}(\bm{k} )} =
({R_1(\bm{k})},{R_2(\bm{k} )},{R_3(\bm{k} )}) $ are real.
The energy dispersions are given by
\begin{eqnarray*}
E_\pm(\bm{k}  ) &=& R_{0}(\bm{k} )  \pm |\bm{R} (\bm{k} )|,
\end{eqnarray*} 
where $|\bm{R}|=\sqrt{R_1^2+R_2^2+R_3^2}$, with the energy gap, $E_g(\bm{k}) $, for each momentum $\bm{k} $ being 
$E_g(\bm{k} )=2| \bm{R} (\bm{k} )|$. 
We have a semiconductor under a condition, 
\begin{eqnarray*}
E_-(\bm{k}_v )\le E_+(\bm{k}_c ),
\end{eqnarray*} 
where $\bm{k}_{v} (\bm{k}_{c})$ are the wave numbers 
in the valence (conduction) band.

In the case of a zero-gap semiconductor, the energy gap 
vanishes at some momentum $\bm{k}_0 $. 
Expanding the Hamiltonian around $\bm{k}_0 $,
we have an effective Hamiltonian (${H}_g\approx {H}  $) as
\begin{eqnarray*} 
H &=&  
 ( X^0\sigma _0  +\bm{X}\cdot  \bm{\sigma } )\delta k_x
+
 (Y^0 \sigma _0   +\bm{Y}\cdot \bm{\sigma } ) \delta k_y ,
\end{eqnarray*} 
where $\delta \bm{k} =\bm{k}-\bm{k}_0$,
$X^0 = \partial _{k_x}R_{0}  |_{\bm{k}_0 }$, 
$Y^0 = \partial _{k_y}R_{0} |_{\bm{k}_0 }$ and the three dimensional vectors $ \widetilde {\bm{X}}  = ({X^1},{X^2},{X^3})$
and
$ \widetilde {\bm{Y}}  = ({Y^1},{Y^2},{Y^3})$ are defined by 
$\bm{X} = \partial _x \bm{R} |_{\bm{k}_0 }$,
$\bm{Y} = \partial _y \bm{R} |_{\bm{k}_0 }$.
The terms that contain $X^0$ or $Y^0$ induce tilting of the Dirac cones, 
while when $(X^0,Y^0)= 0$ the Dirac cones can be anisotropic 
but vertical.

With an effective momentum around the gapless
point, $\bm{p}= \hbar \delta \bm{k}=\widetilde{({p_x},{p_y})}$, 
we have 
\begin{eqnarray} 
{H}  &=& 
\hbar ^{-1} (\sigma_\mu X^\mu   ,\sigma _\mu Y^\mu  ) 
 \bm{p},
\label{effH}
\end{eqnarray} 
where a summation over repeated indices is assumed.%

Now, let us introduced a 
four-dimensional notation to simplify  the calculation. 
For this purpose, we introduce, on top of 
the ``contravariant" four-dimensional vectors
$\widetilde{X}=(X^0,X^1,X^2,X^3)$ and $\widetilde{Y}=(Y^0,Y^1,Y^2,Y^3)$, 
the conjugated (or ``covariant")
vectors $\bar X$ and $\bar Y$
defined as 
$$
\bar X =(X_0,X_1,X_2,X_3)= \widetilde {X} g 
=(-X^0,X^1,X^2,X^3),
$$ 
where
$g={\rm diag}(-1,1,1,1)$ is a metric.
Now we have a simple identity
(see appendix \ref{sec:four}),
\begin{eqnarray} 
( \bar X^\mu \sigma_\mu )(\sigma_\nu Y^\nu )&=&   \bar X Y  \sigma _0 +i\bm{n} \cdot  \bm{\sigma } ,
\nonumber \\
( \bar Y^\mu \sigma_\mu )(\sigma_\nu X^\nu )&=&   \bar  Y X  \sigma _0 -i\bm{n} \cdot  \bm{\sigma } ,\nonumber 
\\
\bm{n} &=&   
\bm{X} \times \bm{Y} + i \bm{\eta},
\label{defn}
\end{eqnarray}
where 
    $
    \bm{\eta} = 
X^0 \bm{Y} 
-Y^0\bm{X}$. 
Note that, while we have $\bar X Y= 
X^\mu Y_\mu 
=\bar Y X $, 
$\bm{n}$ is anti-symmetric against $X \leftrightarrow Y$. 
Its norm becomes  (see Appendix \ref{sec:det})
$$
 \bm{n}^2 = ({\bar X}{X} ) ({\bar Y}{Y} ) - ({\bar Y}{X} ) ({\bar X}{Y} ) \equiv (\hbar c)^4,
$$
where the Fermi velocity of the Dirac fermions, $c$, is defined. When $\bm{n}^2>0 $,  the velocity $c$ is real.

{
  By introducing the covariant notation, the discussion becomes 
transparent. }
For the usual (vertical) Dirac cones, it is known that 
considering a squared Hamiltonian, $H^2$, facilitates the 
analysis.  In the present case of tilted 
cones, this has to be modified.  
We can instead note that it is useful to define 
a Hamiltonian conjugate to Eq.(\ref{effH}) as
\begin{eqnarray*} 
\bar H = \hbar ^{-1} (\bar X^\mu \sigma_\mu   ,\bar Y^\mu \sigma _\mu  ) \bm{p}
= H-2 H_0,
\label{Hbar}
\end{eqnarray*} 
where 
\begin{eqnarray*} 
H_0 &=&  \hbar^{-1} \sigma_0 (X^0 ,Y^0 )\bm{p}.
\end{eqnarray*} 
Now we can consider a product, $\bar H H$, as a 
``contraction" in the present four-dimensional representation.   The expression can be put in a form,
\begin{eqnarray}
\bar H H &=& \hbar ^{-2} \bm{p}  ^\dagger {\cal G} \bm{p}, \nonumber 
\end{eqnarray} 
where ${\cal G}$ is a $4\times 4$ matrix composed of 
$2\times 2$ Pauli matrices and is expressed, with the formula (\ref{defn}), 
as 
\begin{eqnarray*} 
{\cal G} 
&=& 
\mvec
{\bar X^\mu \sigma_\mu  }
{\bar Y^\mu \sigma_\mu  }
(\sigma _\nu X^\nu  , \sigma _\nu Y^\nu )
\\
&=& \mmat
 {{\bar XX}\sigma _0}{{\bar XY} \sigma _0+i\bm{n}\cdot \bm{\sigma }}
{{\bar Y X} \sigma _0-i\bm{n}\cdot \bm{\sigma }}
{{\bar YY} \sigma _0}.
\end{eqnarray*} 
The  determinant of $\cal G $ vanishes, since its rank is two, which can be 
confirmed directly by evaluating the determinant. 
We then have 
\begin{eqnarray}
\bar H H &=& c^{2} \bm{p} ^\dagger \Xi \bm{p} \ \sigma_0,
\\ 
\Xi &=& \frac{1}{(\hbar c)^2}\mmat
{\bar X X}{\bar X Y}
{\bar Y X}{\bar Y Y},
\nonumber 
\end{eqnarray} 
where we have used $[p_x,p_y]=0$ and 
we can note that $\det \Xi =1$. 

From the Schr\"{o}dinger equation, $H\Psi=E\Psi$, 
and Eq.(\ref{Hbar}), we have 
\begin{eqnarray*}
\bar H H \Psi &=& E(H-2H_0)\Psi=(E^2-2EH_0)\Psi,
\end{eqnarray*}
which reduces to 
\begin{eqnarray} 
\sigma_0 \big[
c^2 \bm{p}^\dagger \Xi \bm{p} +2(E/\hbar) (X_0,Y_0)\bm{p}
\big] 
\Psi
&=& E^2\Psi .
\label{eq:comp1}
\end{eqnarray}
By  completing the square, 
we have (detail in Appendix \ref{sec:complete})
a simple, bilinear form,
\begin{equation}
c_r^2 \bm{p}_E^\dagger \Xi \bm{p}_E \sigma _0 \Psi = E^2\Psi, 
%
\label{completedH}
\end{equation} 
where
\begin{eqnarray}
c _r &=& c
\left[
\frac
{{\rm Re\,} \bm{n}^2}
{({\rm Re\,} \bm{n}) ^2
}
\right]^{1/2}
 \equiv \frac c {\cosh q},
\label{cr}
\\
\bm{p} _E &=& \bm{p} +\Delta \bm{p}_E,\nonumber
\\
\Delta \bm{p}_E &=&   E \frac {1}{c^2 \hbar } \Xi ^{-1} \mvec{X^0}{Y^0}.\nonumber
\end{eqnarray} 
This implies the equi-energy contour is an ellipse centered at $\Delta \bm{p}_E  $.
{
  (See Fig.\ref{fig:3} and also Appendices.)
  }
The role of the parameter $q$ appearing in the renormalization factor for the velocity $c$ will 
become apparent when we discuss the relationship between the generalized chiral operator and the conventional chiral operator $\Gamma$ in section IV.

\begin{figure}
\includegraphics[width=6cm]{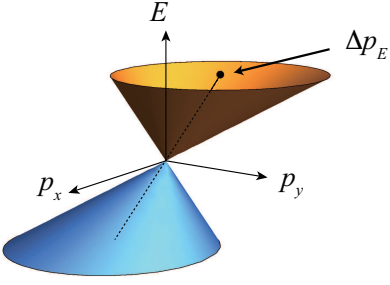}
\caption{A tilted Dirac dispersion and its cross section (an ellipse) with a constant-energy plane. The center of the ellipse is given by $\Delta p_E$.  
 }
\label{fig:3}
\end{figure}

\subsection{Landau levels and the generalized chiral operator}
\label{sec:tlandau}
Having formulated the case in zero magnetic field, 
let us move on to the Landau states 
when we apply an external magnetic field for the tilted Dirac fermion.  In terms of the dynamical momentum $\pi_\mu$ with $ \mu =x,y$,
\begin{eqnarray*}
\pi_\mu &=& p _ \mu -e A_\mu, 
\\
p _ \mu  &=&  -i\hbar \partial _\mu, 
\end{eqnarray*} 
where $e$ is an elementary charge 
and  $A_\mu $ a vector potential which describes a magnetic field 
perpendicular to the two-dimensional system as
\begin{eqnarray*}
B &=& \partial _x A_y- \partial _y A_x.
\end{eqnarray*} 
The dynamical momentum satisfies 
\begin{eqnarray*}
[\pi _x,\pi_y ] &=& i\hbar eB=i(\hbar/\ell_B)^2,
\end{eqnarray*} 
where $\ell_B=\sqrt{\hbar/eB}$ is the magnetic length. 
We may choose $eB>0$ without loss of generality.
With a substitution $\bm{p}\to \bm{\pi} = \bm{p}-e \bm{A}$ 
we have a Hamiltonian,
\begin{eqnarray*}
  H &=& \hbar ^{-1} (\sigma_\mu X^\mu   ,\sigma _\mu Y^\mu  ) \bm{\pi},
\end{eqnarray*} 
and its conjugate,
$$
 \bar H = \hbar ^{-1} (\sigma_\mu {\bar X}^\mu   ,\sigma _\mu {\bar Y}^\mu  ) \bm{\pi} = H -2\sigma_0(X^0,Y^0)\bm{\pi}.
$$
Since $\pi_x$ and $\pi_y$ no longer commute in a magnetic field, we have an extra term proportional to $ \bm{n}\cdot\bm{\sigma}$ for ${\bar H} H$ as 
\begin{eqnarray*}
\bar H H &=& \hbar ^{-2} \bm{\pi}  ^\dagger {\cal G} \bm{\pi} 
\\
&=& 
c ^{2} \bm{\pi}^\dagger \Xi \bm{\pi}  \sigma_0
+ i\hbar ^{-2} \bm{n}\cdot \bm{\sigma } [\pi_x,\pi_y]
\\
&=& 
c ^{2} \bm{\pi}^\dagger \Xi \bm{\pi} \, \sigma_0
- \ell_B^{-2} \bm{n}\cdot \bm{\sigma } .
\end{eqnarray*} 
From the Schr\"{o}dinger equation, $H\Psi = E\Psi$, and the relation above, 
we have  
\begin{equation} 
\left[
c^2 \bm{\pi}^\dagger \Xi \bm{\pi} +\frac{2E}{\hbar} (X_0,Y_0)\bm{\pi}
\right] \sigma^0\Psi
- \ell_B^{-2} \bm{n}\cdot \bm{\sigma } \Psi = E^2\Psi.
\label{eq:comp2}
\end{equation}
We can readily complete the square to arrive at 
\begin{eqnarray*}   
c_r^2 \big[
\bm{\pi}_E ^\dagger  \Xi \bm{\pi}_E
-  (\hbar/\ell_B)^2 \gamma
\big] \Psi = E^2\Psi, 
\end{eqnarray*} 
where $\bm{\pi}_E = \bm{\pi}+ \Delta \bm{p}_E$ with 
$\Delta \bm{p}_E$ and 
$c_r$ defined in Eq.(\ref{cr}).  
An important ingredient is 
the generalized chiral operator $\gamma $, defined 
by
\begin{eqnarray}
\gamma = \frac{\bm{n} \cdot \bm{\sigma }}{(\hbar c)^2},
\label{eq:defGm}
\end{eqnarray} 
which has  eigenvalues $\pm 1$ 
because ${\rm Tr} \gamma =0$, and satisfies
\begin{eqnarray*}
\det \gamma = -\bm{n}^2 /(\hbar c)^4 =-1.
\end{eqnarray*} 
However, the operator is not hermitian in general.

If we denote the right-eigenstates of $\gamma $ as $| \chi_\pm \rangle$  with 
\begin{eqnarray*} 
\gamma |\chi_\pm  \rangle  &=& \pm | \chi_\pm \rangle, 
\end{eqnarray*} 
the wave function is expressed as $\Psi_\pm=|\chi_\pm  \rangle \psi_\pm$.
Then 
the Schr\"{o}dinger equation is reduced to a scalar equation,
\begin{eqnarray*}
 c_r^2 \big[
\bm{\pi}_E ^\dagger  \Xi \bm{\pi}_E
\mp  (\hbar/\ell_B)^2 
\big]\psi_\pm = E^2\psi_\pm.
\end{eqnarray*} 
If we note that the first term, 
$c_r^2 \bm{\pi}_E ^\dagger  \Xi \bm{\pi}_E$, may be mathematically 
regarded, by replacing 
$c_r^2$ with $1/(2m^*)$,  as a 
Hamiltonian for anisotropic 
fermions with
a parabolic dispersion having an effective mass $m^*$ in a magnetic field (Appendix \ref{sec:aniso}) 
, we can introduce
a single-component Landau wave function 
$\psi_n$ that satisfies
\begin{eqnarray*}
\left[\frac {1}{2m^*} \bm{\pi} ^\dagger \Xi\bm{\pi} \right]\psi_n 
&=& 
\hbar \omega_C \left(n+\frac{1}{2}\right)\psi_n ,
\end{eqnarray*}
where the effective cyclotron frequency is 
\begin{eqnarray*} 
 \omega_C &=&  \frac {eB}{m^*}=2c_r^2 eB.
\end{eqnarray*}
The squared energy then has a spectrum, $\hbar \omega _C\big[
(n+ \frac {1}{2} )\mp \frac {1}{2} \big]$, i.e., 
the energy itself has a Dirac Landau level structure, 
\begin{eqnarray*}
  E_n &=& \pm c_r \sqrt{2 \hbar eB {
      n}}, \quad n=0,1,2,\cdots. 
\end{eqnarray*} 
Note that the $n=0$ Landau level is given by the
eigenstate of $\gamma $ with the eigenvalue $+1$. 


\subsection{Generalized chiral symmetry}
\label{sec:echiral}
Let us here discuss the  generalized chiral operator 
$\gamma =\bm{n}\cdot \bm{\sigma }/(\hbar c)^2$ defined in section \ref{sec:tlandau} 
assuming that $c$ is real (i.e., assuming there are Dirac cones).  
Since $\gamma $ is anti-symmetric against $X\rightleftarrows Y$,
we have from Eq.(\ref{defn})
\begin{eqnarray*} 
2i(\hbar c)^2  \gamma &=&\bar x y-\bar y x,
\\
2i(\hbar c)^2  \gamma ^\dagger  &=& x\bar y  -y \bar x ,
\end{eqnarray*}  
where $x=x ^\dagger \equiv \sigma _\mu X^\mu $, $\bar x = \bar x ^\dagger =\bar X^\mu \sigma _\mu $, etc.  
The Hamiltonian can be expressed as  $H=x\pi_x+y\pi_y$, 
so that we obtain
\begin{eqnarray*}
2i(\hbar c)^2 H\gamma &=&
(x\bar x  y-x \bar y x)\pi_x
+(y \bar x y-y \bar y x)\pi_y,
\\
2i(\hbar c)^2 \gamma ^\dagger H &=& 
(x \bar y x-y\bar x x)
\pi_x
+
(x \bar y y-y\bar x y)
\pi_y.
\end{eqnarray*} 
Since $\bar x x=X^\mu X_\mu \sigma_0 $ commutes with $y$
while $\bar y y $ commutes with $x$,
$\gamma$ and $H$ have an anti-commutation relation defined as
\begin{eqnarray*}
\{H, \gamma \}_R &\equiv & H \gamma + \gamma ^\dagger H =0,
\end{eqnarray*} 
which we have called the generalized chiral symmetry.\cite{Kawa11}
Note again that
\begin{eqnarray*}
{\rm Tr}\, \gamma &=& 0,
\ \
\det\gamma = -1,
\\
\gamma ^2 &=& (\gamma ^\dagger ) ^2 = \sigma _0,
\ \
\gamma ^\dagger \ne \gamma .
\end{eqnarray*}  
The generalized chiral symmetry is essential for showing that the zero modes are generally 
eigenstates of the generalized chiral operator. \cite{Kawa11}

\subsection{Robust zero modes}
\label{sec:zero}
Now, let us focus on the zero modes (zero-energy states).
There is a long history of study of the zero modes in 
massless Dirac fermions, notably the well-known work of 
Aharonov and Casher.\cite{ac79,Ludwig94,Kawarabayashi09,Kawa11}
For $E=0$ states, the Schr\"{o}dinger equation $H\Psi=0$
reduces to
\begin{eqnarray*} 
c^2\big[
\bm{\pi} ^\dagger \Xi \bm{\pi} -(\hbar/\ell_B)^2 \gamma \big]\Psi &=& 0.
\end{eqnarray*}  
If we take the eigenstates, $|\chi_+ \rangle   $, of the generalized chiral operator 
with the eigenvalue $+1$ with  $\Psi=|\chi_+ \rangle \psi_+$, $\psi_+$ satisfies
\begin{eqnarray*}
\left[
(\pi_x,\pi_y)  \Xi 
\mvec{\pi_x}{\pi_y}
+i[\pi_x,\pi_y]\right]\psi_+ &=& 0,
\end{eqnarray*} 
since $[\pi_x,\pi_y] = i(\hbar/\ell_B)^2 $. 
The matrix
 $\Xi$, being real symmetric, can be diagonalized with an orthogonal matrix $V_\Xi$ as
\begin{eqnarray}
  {\Xi} &=&  \widetilde {{V}_\Xi} {\rm diag}\, (\xi_1,\xi_2) {V}_\Xi,
  \label{eq:diag}
\end{eqnarray} 
where $\widetilde {{V}_\Xi} {V}_\Xi =\sigma _0, \xi_1>0, \xi_2>0$, and
 $ \xi_1\xi_2=\det \Xi=1$.
Here we have assumed $\det V_\Xi=1$ without loss of generality, since, 
if $\det V_\Xi=-1$, $V_\Xi \sigma _x$ diagonalizes 
$\Xi$ with $\xi_1$ and $\xi_2$ interchanged. 
Then we can define a new momentum,
\begin{eqnarray*}
\bm{\Pi} &=& \mmat{\sqrt{\xi_1}}{0}{0}{\sqrt{\xi_2}} V_\Xi\bm{\pi},  
\end{eqnarray*} 
which preserves the commutator,
\begin{eqnarray*}
[\Pi_1,\Pi_2 ] &=&\sqrt{\xi_1\xi_2} \sum_{i,j} {V_\Xi}_{1i}{V_\Xi}_{2j}[\pi_i,\pi_j]
\\
&=& 
({V_\Xi}_{11}{V_\Xi}_{22}
-
{V_\Xi}_{12}{V_\Xi}_{21})[\pi_x,\pi_y]
\\
&=& \det V_\Xi [\pi_x,\pi_y]
\\
&=& [\pi_x,\pi_y].
\end{eqnarray*} 
The zero-mode equation now reads
\begin{eqnarray*}
D ^\dagger D \psi_+ &=& 0,
\end{eqnarray*}
where
\begin{eqnarray*} 
D &=& \Pi_1 + i\Pi _2.
\end{eqnarray*} 
Since $D ^\dagger D$ is semi-positive definite, we have
\begin{eqnarray*}
D\psi_+ &=& 0.
\end{eqnarray*} 
Noting that this is a first-order differential equation, 
we have an explicit solution (which is given below)  as 
the discussion by
Aharanov-Casher\cite{ac79,Kawa11}. This guarantees the stability of
the zero modes. 
This argument is only possible for a real $c^2$ 
(where the Dirac operator is an elliptic one), 
which explicitly indicates that the index theorem 
for the elliptic operator is indeed relevant.\cite{Kawa11,Kawa12}.

\section{\bf Massive and tilted Dirac fermions}

\subsection{General properties}
Now we come to the massive case in question.  
Our motivation is to clarify the origin of the numerically-observed 
anomalous robustness of 
the split $n=0$ Landau levels for the massive and tilted Dirac fermions.
The generalized chiral operator $\gamma$ introduced in section IIIB can be expressed with the 
normalized vector $\hat{\bm{n} }= \bm{n}/\Delta$, 
which puts Eq.(\ref{eq:defGm}) 
into 
\begin{eqnarray*}
 \gamma &= & \hat{\bm{n}}\cdot \bm{\sigma },
\end{eqnarray*}
where we have introduced the norm of the vector $\bm{n}$,
\begin{eqnarray*}
\Delta &=& \sqrt{\bm{n}^2 }=\sqrt{({\rm Re\,} \bm{n} )^2
-({\rm Im\,} \bm{n})^2}=
 (\hbar c)^2.
\end{eqnarray*} 
Recall in Eq.(\ref{defn})
that real and imaginary parts of $\bm{n}$ are 
given by ${\rm Re\,} \bm{n} =  \bm{X}\times \bm{Y} $ and ${\rm Im\,} \bm{n} =\bm{\eta} =(X_0 \bm{Y}-Y_0 \bm{X})$, so that they are orthogonal with each other 
$[ ({\rm Re\,}  \bm{n} )\cdot ({\rm Im\,}\bm{n}  ) = 0]$.
The conventional chiral operator $\Gamma $ is expressed in a similar form in terms of a real vector $\hat{\bm{n}}_0 =  {\rm Re\,} \bm{n} /\Delta _0$ 
with $\Delta _0 = |{\rm Re\,} \bm{n} |=|\bm{X}\times \bm{Y}  |$ as 
\begin{eqnarray*}
\Gamma &=& \hat{\bm{n}}_0\cdot \bm{\sigma }.
\end{eqnarray*} 

We can then relate $\Gamma $ with 
the generalized chiral operator $\gamma$ as
\begin{eqnarray*}
\gamma &=& 
(\hat{\bm{n}}_0\cdot \bm{\sigma } )
(\hat{\bm{n}}_0 \cdot \bm{\sigma })
(\hat{\bm{n}}\cdot \bm{\sigma })
\\
&=& 
\Gamma 
\left[(\hat{\bm{n}}_0 \cdot \hat{\bm{n}})\sigma _0+ i \bm{\sigma }\cdot (\hat{\bm{n}}_0\times \hat{\bm{n}}) \right]
\\
&=& 
\Gamma 
\left[ (\hat{\bm{n}}_0 \cdot {\rm Re\,} \hat{\bm{n}})\sigma _0-
\bm{\sigma }\cdot (\hat{\bm{n}}_0\times {\rm Im\,} \hat{\bm{n}}) \right]
\\
&=& 
\Gamma \left[  \Delta _0/\Delta -\bm{\sigma }\cdot ({\rm Re\,} \bm{n}
\times {\rm Im\,} {\bm{n} }) / (\Delta _0 \Delta )\right],
\end{eqnarray*} 
where we have inserted $\Gamma^2=1$ in the first line,  used 
a formula above Eq.(A1) in the second line 
and the fact that $\hat{\bm{n}}_0 \perp {\rm Im\,} \hat{\bm{n}}$ in the third line.  
Since 
$
|{\rm Re\,} \bm{n} \times {\rm Im\,}{\bm{n}} | =
|{\rm Re\,} \bm{n} ||{\rm Im\,}{\bm{n}} | =\Delta _0
\sqrt{\Delta _0^2-\Delta ^2}
$, we end up with a compact expresseion,
\begin{equation*}
\gamma = \Gamma (\cosh q - {\bm{\tau}\cdot \bm{\sigma }   }\sinh q)
=\Gamma e^{-q \bm{\tau}\cdot \bm{\sigma }  },
\end{equation*}
where
the parameter $q$ is defined in  Eq.(\ref{cr})
or equivalently $\tanh q =   \sqrt{\Delta _0^2-\Delta ^2}/\Delta = |\bm{\eta}|/|\bm{X} \times \bm{Y}|$, and the unit vector
$\bm{\tau}$ is given by
\[ 
\bm{\tau} = \frac{ {\rm Re\,}\bm{n}\times {\rm Im\,} \bm{n}}{|{\rm Re\,} \bm{n} \times {\rm Im\,} \bm{n}|}
= \frac{(\bm{X}\times \bm{Y})\times \bm{\eta}}{|(\bm{X}\times \bm{Y})\times \bm{\eta}|}. 
\]
Note that the parameter $q$ is real as long as $\Delta^2 \ge 0$, {
  which is equivalent to the ellipticity of the Hamiltonian (\ref{effH}) where the index theorem is relevant.
}

We can also note that $\{\Gamma , \bm{\tau}\cdot \bm{\sigma }  \}=0$, 
since $\bm{\tau}$ is normal to $\bm{X} \times \bm{Y}$, 
and we have a suggestive representation, 
\begin{eqnarray*}
 \gamma &=& 
\Gamma e^{-q\bm{\tau}\cdot \bm{\sigma}} 
=
e^{q\bm{\tau}\cdot \bm{\sigma}} \Gamma 
=
e^{q\bm{\tau}\cdot \bm{\sigma}/2}
  \Gamma \
e^{-q\bm{\tau}\cdot \bm{\sigma}/2}.
\end{eqnarray*} 
This immediately implies that the eigenstates $
|\pm \rangle$ of the conventional (hermitian) 
chiral operator $\Gamma$ 
(with $\Gamma |\pm \rangle = \pm |\pm \rangle$)
 can be related to the right-eigenstates $|\chi_\pm \rangle $ of 
the generalized (non-hermitian) chiral operator 
as 
\begin{eqnarray*} 
 |\chi_\pm \rangle &=&  \frac{1}{\sqrt{\cosh q}} e^{q\bm{\tau}\cdot \bm{\sigma}/2} |\pm \rangle.
\end{eqnarray*} 
The normalization factor $1/\sqrt{\cosh q}$ is introduced, since 
$\langle + | \exp(q(\bm{\tau}\cdot \bm{\sigma}))| + \rangle =\langle - | \exp(q(\bm{\tau}\cdot \bm{\sigma}))| - \rangle= \cosh q$. 
On the other hand, we can readily verify a relation, 
\begin{equation}
 \gamma^\dagger  \Gamma \gamma = \Gamma,
 \label{ch-ch-relation}
\end{equation}
which guarantees that 
$$
 \langle \chi_+ | \Gamma | \chi_- \rangle = \langle \chi_- | \Gamma | \chi_+ \rangle = 0 .
$$
The diagonal matrix elements are evaluated as
$$
 \langle \chi_\pm | \Gamma | \chi_\pm \rangle = \pm \frac{1}{\cosh q} = \pm \frac{\Delta}{|\bm{X}\times \bm{Y}|}.
$$

\subsection{Symmetry breaking and robust zero modes
}

The relations obtained above are useful in considering the effects of the mass term (i.e., a staggered field $\propto \Gamma$), which breaks the 
generalized chiral symmetry into {
  $\{H,\gamma \}_R\neq 0$ for the Hamiltonian $H$}. 
For the vertical Dirac cones,{
  \cite{Haldane88,EPLSUSY09,Fuchs}} the effect of the staggered potential is rather trivial, since 
the states in the $n=0$ Landau level are also eigenstates of the chiral operator $\Gamma$, with their energies simply shifted 
according to their eigenvalues of $\Gamma$.
By sharp contrast,  tilted Dirac cones have 
the states in the $n=0$ Landau level that 
reside on both of the sub-lattices, 
and are not the eigenstates of $\Gamma$. This is why 
the effects of the staggered potential becomes 
nontrivial for tilted cone.
We now employ the representation of $\Gamma$ in terms 
of the generalized chiral bases to explore 
the effects of the staggered potential on the $n=0$ Landau level.  Essentially, we shall show that    
the states in the $n=0$ Landau level remain the 
eigenstates of $\gamma$ even in the presence of the staggered potential.

For a typical source of mass gap, 
we can again introduce a chiral symmetry breaking term  $m c^2\Gamma$ in the Hamiltonian as
\begin{eqnarray*}  
 H(m) &=&  H + m c^2\Gamma.
\end{eqnarray*} 
For the massless, tilted cones, we have shown that 
it is usuful to consider 
$\bar H H$.  
Let us extend this argument to the massive case 
by considering $\bar H(m) H(m)$.  Amazingly, 
we can simplify this into
\begin{eqnarray*}
\bar H(m) H(m) &=& (\bar H+ mc^2 \Gamma )(H+m c^2\Gamma ) 
\\
&=& \bar H H+ m ^2 c^4,
\end{eqnarray*}
where cross terms between $\bar H(m)$ and $H(m)$ 
vanish because the unperturbed 
Hamiltonian without tilting, $H_C=H-H_0$, is chiral symmetric with $\{H_C,\Gamma \} = 0$. 
Now, following the case without tilting, let us assume 
that the 
$n=0$ Landau state to be $\Psi^m = |\chi_+ \rangle \psi_+^m$.
Then the Schr\"odinger equation, $H(m)\Psi^m =E\Psi^m$, implies
\begin{eqnarray*}
\bar H(m) H(m) \Psi^m &=&  (E^2-2EH_0)\Psi^m
\\
&=&  
  {
  }
  c^2\big[\bm{\pi} ^\dagger \Xi \bm{\pi}  -(\hbar /\ell_B)^2 \gamma +m^2c^2 \big]\Psi^m,
\end{eqnarray*} 
which leads to 
\begin{equation}
c_r^2\big[\bm{\pi}_E ^\dagger \Xi \bm{\pi}_E  -(\hbar /\ell_B)^2  + m^2 c^2\big]
\psi_+^m
= E^2 \psi_+^m.
\label{eq-sq}
\end{equation} 
It is clear from this equation that the symmetry breaking term $mc^2\Gamma$ indeed opens 
a gap $\pm mcc_r$ in the absence of a magnetic field.
We can cast this into
\begin{eqnarray*}
c_r^2( D_E ^\dagger D_E +m^2c^2)\psi_+^m &=& E^2\psi_+^m,
\label{DEdaggerDE}
\end{eqnarray*} 
where 
\begin{eqnarray*}
D_E &=& \Pi_{1,E}+ i \Pi_{2,E},
\\
\bm{\Pi}_E &=& 
\mmat{\sqrt{\xi_1}}{0}{0}{\sqrt{\xi_2}}{V}_\Xi \bm{\pi} _E
\end{eqnarray*} 
with $\xi_1,\xi_2$ given in Eq.(\ref{eq:diag}).  
{
  Then $D_E ^\dagger D_E$ is semipositive definite, and 
the wave function in the $n=0$ Landau level 
is specified  by 
\begin{eqnarray*}
D_E\psi_+ ^m &=& 0, 
\end{eqnarray*}
which has an energy
\begin{eqnarray*}
E &=& mc  c_r= mc^2/\cosh q,
\end{eqnarray*}  
where we have used Eq.(\ref{cr}), and 
chosen the positive sign for the energy 
since it should tend to $+mc^2$ when the tilting 
become zero {
  (See Fig.\ref{fig:4})}.  
As far as the velocity $c$ is real, 
$D$ and $D_E$
(also $\bm{\pi}$ and $\bm{\pi}_E $)
are simply related through a shift in the momentum by $\Delta\bm{p}_E$, 
which indicates 
$\psi_+$  and $\psi_+^m$ are also 
related via a gauge transformation,
\begin{equation}
\psi_+^m = e^{-i \Delta\bm{p}_E \cdot \bm{r} / \hbar }\psi_+.
\label{gauge-trans}
\end{equation} 

\begin{figure}
\includegraphics[width=8cm]{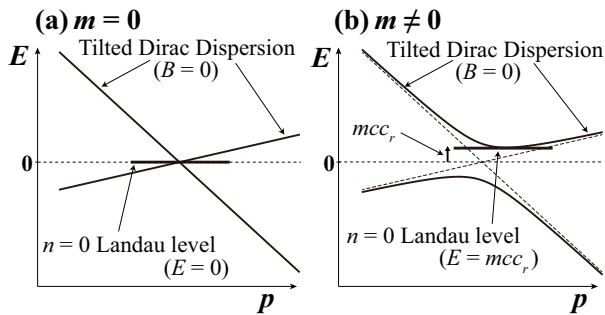}
\caption{The $n=0$ Landau level (horizontal lines) for tilted Dirac fermions. While its energy is zero for the massless ($m=0$) case (a),  it is shifted to $+mcc_r$ in the 
presence of the mass term $mc^2\Gamma$ ($m\neq 0$) when $eB >0$  (b).  }
\label{fig:4}
\end{figure}

  Since ${[}D_E,D_E ^\dagger {]}=-2i {[}\Pi_1,\Pi_2 {]}=2e B\hbar$,
we have  Landau levels for generic Dirac fermions with mass and tilting  as
\begin{eqnarray*}
  E_n &=&
  \mcr{
    \mchss
      { mcc_r, }{ n=0}
      {\pm c_r \sqrt{2e B \hbar n+ m^2 c^2},}{ \quad\quad n=1,2,\cdots.}
      }
\end{eqnarray*}


\mcr{
  This is a condensed matter realization of
  the anomaly\cite{PhysRevLett.53.2449,PhysRevD.29.2375}.
  As for a Dirac cone of a specific
  lattice model, the chiral operator $\Gamma  $ and the 
  mass term are determined in a model dependent way for each valley.
  See for example,
  Eq.(\ref{eq:exValley}).
  It implies the  $n=0$ Landau level is not valley degenerated
  in two dimensional semiconductors with small gap.
  }
To grasp the role of the generalized chiral
{
  operator}
and the relation (\ref{ch-ch-relation}) more explicitly,  let us  
write the wave function $\psi$ in the chiral basis as $\Psi^m= |\chi_+\rangle \psi_+^m + |\chi_-\rangle \psi_-^m$. 
Then the Schr\"{o}dinger equation, $H(m) \Psi^m=E\Psi^m$, becomes 
\begin{eqnarray*}
 \lefteqn{\left(\begin{array}{cc}
 \langle \chi_+ | mc^2\Gamma | \chi_+\rangle & \langle \chi_+ | H | \chi_-\rangle \\
 \langle {\chi_-} | H | \chi_+\rangle & \langle \chi_- | mc^2\Gamma |\chi_-\rangle
 \end{array}\right)
 \left(\begin{array}{c}
 \psi_+^m \\
 \psi_-^m
 \end{array}\right) } \\ 
 & & \quad \quad \hspace{3cm}= E
 \left(\begin{array}{cc}
 1 & \beta \\
 \beta^* & 1
 \end{array}\right)
 \left(\begin{array}{c}
 \psi_+^m \\
 \psi_-^m
 \end{array}\right),
\end{eqnarray*}
where $\beta = \langle \chi_+ |\chi_-\rangle$. Due to the generalized chiral symmetry, $H$ 
appears only in the off-diagonal elements, while  
the relation (\ref{ch-ch-relation}) guarantees that $\Gamma$ appears only in the diagonal elements.  
From the explicit form of the matrix elements for 
$\Gamma$, the equation is simplified to 
\begin{equation}
 \left(\begin{array}{cc}
 mcc_r & \bm{\alpha}\cdot \bm{\pi}_E\\
 \bm{\alpha}^*\cdot \bm{\pi}_E & - m cc_r
 \end{array}\right)
 \left(\begin{array}{c}
 \psi_+^m \\
 \psi_-^m
 \end{array}\right) = E
 \left(\begin{array}{c}
 \psi_+^m \\
 \psi_-^m
 \end{array}\right),
 \label{chiral_basis}
\end{equation}
with 
$$
 \widetilde{\bm{\alpha}} \equiv (\alpha_X, \alpha_Y)\\
 = \hbar^{-1}(\langle \chi_+ | X^\mu \sigma_\mu |\chi_-\rangle,
\langle \chi_+ | Y^\mu \sigma_\mu  |\chi_-\rangle)
$$
(see Appendix \ref{equivalence}).
Then we find 
that the normalizable wave functions for $eB>0$ at $E = \pm mcc_r$ should have \cite{Kawa11}
$$\psi_-^m = 0, \quad \bm{\alpha}^*\cdot \bm{\pi}_E \psi_+^m = 0, 
$$
which indicates that the eigenstates at the bottom of the upper band ($E=mcc_r$) are indeed the 
eigenstate of $\gamma$ with the eigenvalue $+1$ ($\Psi^m = |\chi_+\rangle \psi_+^m$), because $\psi_-^m=0$. 
{
  Namely, the generalized chiral operator 
continues to commute with the Hamiltonian, within the $n=0$ Landau subspace, even for massive Dirac fermions.}

In other words, for tilted Dirac fermions, the wave functions of the $n=0$ Landau levels for massless ($m=0$)  and those for 
massive ($m\neq 0$) fermions are related through the gauge transformation (\ref{gauge-trans}).
We can therefore conclude  that the robustness of the  $n=0$ Landau level at 
$E=0$ against disorder that respects the 
generalized chiral symmetry persists to the cases where its energy is shifted to $E=mcc_r$ by the mass term $mc^2\Gamma$. 

In the tight-binding lattice model discussed in section II, we have two valleys, for which the sign of the symmetry-breaking term $mc^2\Gamma$ is 
opposite. The sign of the energy shift is therefore opposite for these two Dirac cones in the lattice model, which is actually seen as 
the split zero modes shown in Fig. \ref{fig:1}. 
We show in Fig. \ref{fig:5} the energies of the split $n=0$ Landau levels obtained for the tight-binding lattice model (\ref{lattice_model}) as 
a function of  $mc^2$. They excellently 
agree with the analytical formula, 
$\pm mcc_r$, given by the effective theory.

\begin{figure}
\includegraphics[width=9cm]{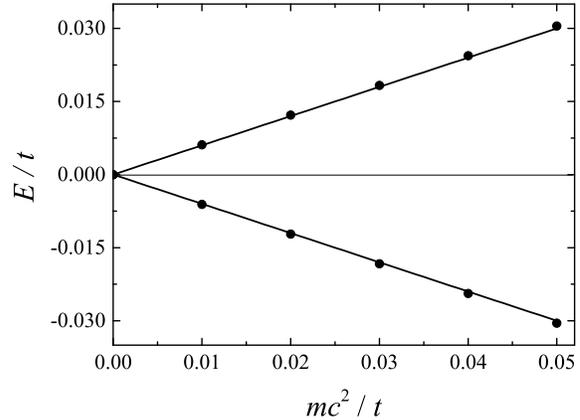}
\caption{The energies for the split $n=0$ Landau levels as a function of the mass term $mc^2$ for the tight-binding 
lattice model, Eq.(\ref{lattice_model}) (solid circles). Solid 
lines represent the energies expected from the effective theory, $\pm mcc_r = \pm mc^2\sqrt{1-4(t'/t)^2} 
(= \pm 0.6mc^2$ here). 
 }
\label{fig:5}
\end{figure}

\section{Summary}
We have investigated the robustness of the zero modes for massive and tilted Dirac fermions in a magnetic field.
It is demonstrated numerically that the anomalous robustness of zero modes against disorder in gauge degrees of freedom 
is preserved for a massive and tilted Dirac fermions.  Notably, for the massive fermions, 
the robustness appears even in the case of the short-range disorder, 
in contrast with the case of massless Dirac fermions. 
We have also presented a general formulation for the generic two-dimensional massless and massive Dirac fermions in which 
a simple algebraic transformation between the generalized chiral operator and the conventional chiral operator has been obtained.
Based on the low-energy effective theory, we have explicitly discussed the applicability of the argument by Aharonov and Casher to 
show the robust zero modes of the massive and tilted Dirac fermions, where the wave function of the 
$n=0$ Landau level for the massive case is  related to that for the massless case through a gauge transformation. 
The present numerical and analytical results for tilted Dirac fermions, where the chiral symmetry and 
the sub-lattice symmetry are distinguished,  
clearly suggest that the generalized chiral symmetry, rather than the sub-lattice symmetry, is indeed a key ingredient for the robust zero modes for 
the generic Dirac fermions in two dimensions.

\begin{acknowledgments}
We thank Y. Ono and T. Morimoto
for useful discussions and comments.
This work was supported in part by JSPS Grant Numbers 26247064, 25610101(YH), 25107005(HA)
 and 25610101 from MEXT(YH,HA).
\end{acknowledgments}

\appendix
\section{Four dimensional notation}
\label{sec:four}
Let us {
  elaborate } the general formulation 
for four-dimensional real vector, 
$X=^{t}\!\!(X_0,X_1,X_2,X_3)$ with a metric
$g = {\rm diag}\, (-1,1,1,1)$, 
which defines
\begin{eqnarray*} 
\bar X &=& (X^0,X^1,X^2,X^3)=
^t\!\! {X} g=(-X_0,X_1,X_2,X_3)
\end{eqnarray*}  
An inner product of the two 4-vectors $\bar X$ and $Y$ is
expressed as
\begin{eqnarray*}
\bar X Y &=& \bar Y X =
 X_\mu Y^\mu =- X^0 Y^0+ \bm{X} \cdot \bm{Y}. 
\end{eqnarray*} 
For example, the norm of the four vector $\bar X X$ is given as
\begin{eqnarray*}
\bar  X X &=& X^\mu X_\mu =|\bm{X} |^2-X_0^2.
\end{eqnarray*} 
Noting that for three dimensional vectors $\bm{X} $ and $\bm{Y} $, one has
$(\bm{X}\cdot \bm{\sigma })(\bm{Y}\cdot \bm{\sigma })=
(\bm{X}\cdot \bm{Y}) \sigma _0+ i (\bm{X}\times \bm{Y})\cdot \bm{\sigma }  $, 
we have a simple
formula,
\begin{eqnarray}
(\bar X^\mu \sigma _\mu )(\sigma _\mu Y^\mu )
&=&  (-X^0 \sigma _0+ \bm{X}\cdot \bm{\sigma }  )
(Y^0 \sigma _0+ \bm{Y}\cdot \bm{\sigma }  ) 
\nonumber \\
&=&   \bar X Y  \sigma _0 
+i\bm{n} (X,Y)\cdot  \bm{\sigma }, 
\label{ap:eDirac}
\end{eqnarray}
where
\begin{eqnarray} 
\bm{n}(X,Y)  &=& 
\bm{X} \times \bm{Y} + i \bm{\eta} (X,Y),
\\
\bm{\eta} (X,Y)&=& 
X^0 \bm{Y} 
-\bm{X}Y^0  .
\end{eqnarray} 
Note that 
$\bar X Y=X \bar Y\equiv 
X^\mu Y_\mu =-X^0Y^0+ \bm{X}\cdot \bm{Y}  $
is symmetric, while
$\bm{n}(X,Y) $ is anti-symmetric when one 
exchanges $X$ and $Y$. 

Also, noting that
\begin{eqnarray*}
\det A^\mu \sigma  _\mu &=& 
\det
\mmat
{A^0+A^3}{A^1-iA^2}
{A^1+iA^2}{A^0-A^3}
\\
&=& (A^0)^2-|\bm{A} |^2
= -\bar A A,
\end{eqnarray*} 
%
we have, by  defining $\sigma _{\bm{A} } \equiv {\bm{A}}\cdot \bm{\sigma }/|\bm{A}|$,
\begin{eqnarray*}
A^\mu \sigma _\mu &=& A^0 \sigma _0 +  \bm{A} \cdot \bm{\sigma } 
\\
&=& 
\sqrt{\bar A A} e^{ \phi _A \sigma _{\bm{A} }} \sigma _{\bm{A} },
\end{eqnarray*}
where $e^{\phi_A \sigma_{\bm{A}}  } = \sigma _0  \cosh\phi_A 
+ \sigma _{\bm{A} }\sinh \phi_A$ with $\cosh \phi _A = |\bm{A} |/\sqrt{\bar A A}$ and 
$\sinh \phi_A = A_0/\sqrt{\bar A A}$. 

\section{Determinant of $\Xi$}
\label{sec:det}
Let us here evaluate the determinant in the discussion
as
\begin{eqnarray*} 
  (c \hbar )^4 &=& \lmat
{\bar X X}
{\bar X Y}
{\bar Y X}
{\bar Y Y}
\\
&=& 
 (|\bm{X}|^2-X_0^2)(|\bm{Y}|^2-Y_0^2)
-\big(\bm{X}\cdot \bm{Y}-X_0Y_0)^2  
\\
&=& 
|\bm{X} \times \bm{Y} |^2 -|X^0 \bm{Y} - \bm{X} Y^0|^2
\\
&=&  {\rm Re\,} (\bm{n}\cdot \bm{n})
\\
&=&  \bm{n}^2,
\end{eqnarray*} 
where
${\rm Im\,} (\bm{n}\cdot \bm{n})=  2(\bm{X}\times \bm{Y}  )\cdot(X^0 \bm{Y}-\bm{X}Y^0  ) =0 $.

It is also evaluated by the expansion of the minors as
\begin{eqnarray*}
\lmat
{\bar X X}{\bar X Y}
{\bar Y X}{\bar Y Y}
&=& 
\det \bigg[
\left(
\begin{array}{cccc}
-X^0 & X_x & X_y & X_z \\
-Y^0 & Y_x & Y_y & Y_z 
\end{array}
\right)
\left(
\begin{array}{cc}
X^0 & Y^0 \\
X_x & Y_x \\
X_y & Y_y \\
X_z & Y_z 
\end{array}
\right)
\bigg]
\\
&=& 
\lmat {-X^0 }{ X_x}{-Y^0 }{ Y_x }
\lmat {X^0 }{ Y^0 }{X_x }{ Y_x } \\
&&+\lmat{-X^0 }{ X_y}{-Y^0 }{ Y_y }
\lmat{X^0 }{ Y^0 }{X_y }{ Y_y }\\
&&+\lmat{-X^0 }{ X_z}{-Y^0 }{ Y_z }
\lmat{X^0 }{ Y^0 }{X_z }{ Y_z }\\
&& + \det \bigg[
\left(
\begin{array}{ccc}
 X_x & X_y & X_z \\
 Y_x & Y_y & Y_z 
\end{array}
\right)
\left(
\begin{array}{cc}
X_x & Y_x \\
X_y & Y_y \\
X_z & Y_z 
\end{array}
\right)\bigg]
\\
&=& -(X^0 Y_x-Y^0 X_x)^2 -(X^0 Y_y-Y^0 X_y)^2\\
&&-(X^0 Y_z-Y^0 X_z)^2+|\bm{X} \times \bm{Y} |^2\\
&=& 
|\bm{X} \times \bm{Y} |^2
-|X^0 \bm{Y}- \bm{X} Y^0 |^2
\\
&=&  {\rm Re\,} \bm{n}^2
=   \bm{n}^2
\end{eqnarray*} 

\section{Completing the square}
\label{sec:complete}
Here let us show details for deriving Eq.(\ref{completedH}) 
by completing the square. We start with 
\begin{eqnarray*}
 (X^0,Y^0)&\Xi ^{-1} &
\mvec{X^0}{Y^0} 
{(\hbar c)^2} 
\\
&=&  
(X^0,Y^0)
\mmat
{\bar YY}{-\bar XY}
{-\bar YX}{\bar XX}
\mvec{X^0}{Y^0} 
\\
&=&  |X^0 \bm{Y}  -Y^0 \bm{X} |^2 \\
&=& |\bm{\eta} (X,Y)|^2
\\
&=&   ({\rm Im\,} {\bm{n}} )^2.
\end{eqnarray*} 
Then we have
\begin{eqnarray*}
&&c ^2 \bm{p} ^\dagger \Xi \bm{p}  + 2(E/\hbar )(X^0,Y^0) \bm{p} 
\\
&=& 
\left[c \bm{p} ^\dagger  +\frac {E}{c\hbar}(X^0,Y^0)\Xi ^{-1} \right]
\Xi 
\left[c \bm{p} +\frac {E}{c\hbar}\Xi ^{-1}\mvec{X^0}{Y^0} \right]
\\
&& - 
\frac {E^2}{(c\hbar)^2}(X^0,Y^0)\Xi ^{-1} 
\mvec{X^0}{Y^0} 
\\
&=& 
c^2 \bm{p}_E ^\dagger  \Xi \bm{p} _E ^\dagger -
\frac{({\rm Im\,} \bm{n} )^2}{(c \hbar )^4} E^2,
\end{eqnarray*} 
where
\begin{eqnarray*}
\bm{p} _E &=& \bm{p} +\Delta \bm{p}_E,
\\
\Delta \bm{p}_E &=&   E \frac {1}{c^2 \hbar } \Xi ^{-1} \mvec{X^0}{Y^0}.
\end{eqnarray*} 
{Also note that
\begin{eqnarray*}
1+\frac{({\rm Im\,} \bm{n} )^2}{(c \hbar )^4} 
&=& 
\frac{\bm{n}^2 +({\rm Im\,} \bm{n} )^2}{\bm{n}^2} 
=
\frac{({\rm Re\,} \bm{n}) ^2}{{\rm Re\,} (\bm{n}^2)} 
\\
&=& {(\cosh q)^2}. 
\end{eqnarray*} 
}

\section{Three-dimensional representation}
In the main text, we have given a compact treatment of the 
tilted Dirac cone physics with a four-dimensional representation.  
Here, let us show that how a three-dimensional treatment 
is feasible but cumbersome.  
The Schr\"{o}dinger equation for the 2-component spinor $\Psi$ is given as
\begin{eqnarray*}
H \Psi 
&=& 
\big[
\hbar ^{-1} \sigma _0 (X^0,Y^0)\bm{p}\, 
+
{H_C^0}\big] \Psi
=
 E\Psi, 
\end{eqnarray*}
where 
$\widetilde {\bm{p} } = ({p_x},{p_y})$ and 
$H_C^0 = \hbar ^{-1} (\bm{X}\cdot\bm{\sigma},\bm{Y}\cdot\bm{\sigma})\bm{p}$.
The equation is written as
$
{H_C^0}
 \Psi 
= 
(E-z)\Psi
$
with
\begin{eqnarray*} 
z &=& \hbar ^{-1} (X^0,Y^0) \bm{p}.
\end{eqnarray*} 
 Using it twice, one has 
\begin{eqnarray*}
({H_C^0})^2 \Psi &=& (E-z)^2\Psi.
\end{eqnarray*} 
Since
$(H_C^0)^2=  [ c_0^2 \bm{p}   ^\dagger \Xi_0 \bm{p}] \, \sigma _0 \propto \sigma_0$
\footnote{ 
Using a well known formula, we have
$ (H_C^0)^2
=    \hbar  ^{-2} {[}\bm{\sigma }\cdot (\bm{X}, \bm{Y} )\bm{p} ] ^ 2
= 
\hbar  ^{-2}{[}
(\bm{\sigma }\cdot \bm{X})
 p_x
+
(\bm{\sigma }\cdot \bm{Y})
 p_y
\big{]}^2
=  
\hbar  ^{-2}(\bm{X} \cdot \bm{X} p_x^2
+
2\bm{X} \cdot \bm{Y} p_xp_y
+
\bm{Y} \cdot \bm{Y} p_y^2) \sigma _0 
=   | (\bm{X},\bm{Y})  \bm{p} |^2  \sigma _0
=  [ c_0^2 \bm{p}   ^\dagger \Xi_0 \bm{p}] \, \sigma _0
$.
}, 
we have a scalar equation  for $\Psi$, 
\begin{eqnarray}
\hbar ^{-2}   [(\bm{X},\bm{Y}) \bm{p}]^2 =
c_0^2 \bm{p}   ^\dagger \Xi_0 \bm{p}=
(E-z)^2,
\label{eq:tcone}
\end{eqnarray}
where 
\footnote{ 
$
\det {\Xi}_0 = \frac {1}{(c_0\hbar)^4} 
\big(|\bm{X} |^2|\bm{Y} |^2-(\bm{X}\cdot \bm{Y}  )^2\big)=
 \frac {1}{(c_0\hbar)^4} |\bm{X}\times \bm{Y}|^2=1
$
}
\begin{eqnarray}
\Xi_0 &=&  \frac {1}{(\hbar c_0)^2} 
\mmat
{\bm{X}\cdot \bm{X} }{\bm{X}\cdot \bm{Y}  }
{\bm{X}\cdot \bm{Y}  }{\bm{Y} \cdot \bm{Y} },
\label{eq:Xi0U}
\\
 c_0^2  &=&   | \bm{X} \times \bm{Y} |/\hbar^2.
\label{eq:defc0}
\end{eqnarray} 
The `` light velocity''  $c_0$ is so chosen that $\det \Xi_0=1$.

Geometrically (see Fig.\ref{fig:cones}), a constant energy curve
$E(p_x,p_y)=$const. 
in $(p_x,p_y,\xi)$ space 
is given by
the intersection of the cone  and the plane
\begin{eqnarray*} 
\xi ^2 = c_0^2 \bm{p} ^\dagger \Xi_0 \bm{p}, 
\\
(X^0/\hbar ) p_x + (Y^0/\hbar )  p_y+\xi  =  E,
\end{eqnarray*} 
which can be  a parabola, an ellipse, hyperbola or a point. 
Any intersection of the cone and the plane is an ellipse if the slope of the
plane does not exceed that of the cone, 
which guarantees that the energy dispersion is given by the 
Dirac cone. 

\begin{figure}
\includegraphics[scale=0.3]{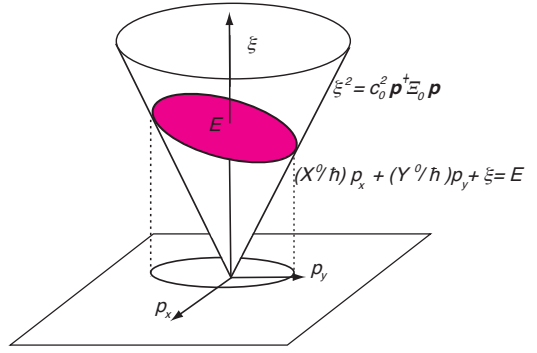}
\caption{
  Geometrical meaning of the tilted Dirac cones in $(p_x,p_y,\xi)$ space.
\label{fig:cones}
}
\end{figure}

When the Dirac cone is not tilted, that is  $X^0=Y^0$=0, 
the energy dispersion is given by $E=z$.
Since ${\Xi}_0 $ is a real symmetric matrix with 
${\rm Tr\, }{\Xi}_0>0 $, 
it is diagonalized by the orthogonal matrix ${V} $ as
\begin{eqnarray*}
{\Xi}_0 &=&  {V}  ^\dagger {\rm diag}\, (\xi^0_1,\xi^0_2) {V}, 
\end{eqnarray*} 
where $  \xi^0_1>0, \xi^0_2>0$ and $ \xi^0_1\xi^0_2=\det \Xi_0=1$.
Now we have
\begin{eqnarray*} 
E &=& \pm c_0 \bar P,
\end{eqnarray*}
where
$ \bar P = \sqrt{\xi^0_1 P_1^2+\xi^0_2 P_2^2}
$,
$\bm{P}={V} \bm{p} $ and $c_0$ is the Dirac fermion 
 velocity without tilting.

For the tilted case with finite $X^0$ or/and $Y^0\neq 0$, 
we need to complete the square by rewriting 
Eq.(\ref{eq:tcone}).
Here let us complete the square in  
Eq.(\ref{eq:tcone}). If we expand the right-hand side as
\begin{eqnarray*}
&&
c_0 ^2 
\bm{p} ^\dagger
\Xi_0
\bm{p}
= 
\hbar ^{-2} 
\bm{p} ^\dagger
\mvec{\bm{X}}{ \bm{Y}}
(\bm{X},\bm{Y})  \bm{p}
\\
&=& 
\bigg[ E-
\big[
(\frac{X_0}{\hbar}) p_x
+
(\frac{Y_0}{\hbar}) p_y 
\big]
\bigg]^2
\\
&=& \bigg[
E^2 -2E\hbar ^{-1} (X^0,Y^0) \bm{p} 
 +\hbar ^{-2} \bm{p}^\dagger 
\mvec
{X^0}{Y^0}
(X^0, Y^0)
\bm{p} \bigg],
\end{eqnarray*}
we have 
\begin{eqnarray*}
&&
\left[
c^2 \bm{p}  ^\dagger  \Xi  \bm{p}  
+2\left(\frac{E}{\hbar}\right)(X^0,Y^0) \bm{p} 
\right]\Psi
=  E^2\Psi,
\end{eqnarray*}
where
\begin{eqnarray*} 
 \Xi &=& \frac {1}{(\hbar c)^2}  
\mmat
{-X^0 X^0 +\bm{X}\cdot \bm{X}  }
{-X^0 Y^0 +\bm{X}\cdot \bm{Y}}
{-X^0 Y^0 +\bm{X}\cdot \bm{Y}}
{-Y^0 Y^0 +\bm{Y}\cdot \bm{Y}}
\\
 &=& 
 \frac {1}{(\hbar c)^2}  
\mmat
{\bar X X}{\bar X Y}
{\bar X Y}{\bar Y Y}.
\end{eqnarray*}
The equation coincides with Eq.(\ref{eq:comp1}) 
in the four-dimensional notation in the text.

Although complicated, one can  perform a similar process
with a magnetic field as
\begin{eqnarray*} 
H &=& \hbar ^{-1} \big[
\sigma _\mu (X^\mu ,Y^\mu ) \bm{\pi} \big] 
= H_0+H_C ,
\\
H_0 &=&  \hbar ^{-1}\sigma _0 (X^0,Y^0)\bm{\pi} ,
\\
H_C &=& 
 \hbar^{-1} (\bm{X}\cdot \bm{\sigma},\bm{Y}  \cdot \bm{\sigma})
\bm{\pi},
\end{eqnarray*}  
where  $\bm{\pi}=\bm{p} -e\bm{A}$ is the dynamical momentum. The Schr\"{o}dinger equation reads
\begin{eqnarray*} 
H_C \Psi &=& (E-Z)\Psi
\end{eqnarray*}
with 
$Z = 
 (X^0/\hbar) \pi_x+(Y^0/\hbar)\pi_y
$.
Using the relation
\begin{eqnarray*}
[H_C,Z] &=& 
[\hbar ^{-1} \big(
(\bm{X}\cdot\bm{\sigma})
\pi_x
+
(\bm{Y}\cdot\bm{\sigma })
\pi_y
\big)
 ), Z]
\\
&=& 
\hbar^{-2} (
 Y^0 \bm{X} 
-
 X^0 \bm{Y} )\cdot \bm{\sigma } 
)
 [\pi_x,\pi_y]
\\
&=& -i \ell_B^{-2}  ( {\rm Im\,} \bm{n}) \cdot \bm{\sigma } 
\end{eqnarray*}
and
\begin{eqnarray*}
H_C^2 &=&  \hbar ^{-2} \left[
(\bm{X}\cdot\bm{\sigma})
\pi_x
+
(\bm{Y}\cdot\bm{\sigma })
\pi_y
]\right]^2 
\\
&=& 
(\bm{\pi}  ^\dagger c_0^2 \Xi_0 \bm{\pi} ) \sigma _0
+ i \hbar ^{-2}  (\bm{X} \times \bm{Y} )\cdot \bm{\sigma } [\pi_x,\pi_y]
\\
&=& 
(\bm{\pi}  ^\dagger c_0^2\Xi_0 \bm{\pi} ) \sigma _0
-\ell_B ^{-2}  ({\rm Re\,} \bm{n} ) \cdot \bm{\sigma } ,
\end{eqnarray*} 
we have
\begin{eqnarray*}
H_C^2\Psi &=& H_C\left[(E-Z)\Psi\right]
\\
&=& 
\bigg[ (E-Z) H_C  +[H_C,E-Z])\bigg]\Psi
\\
&=& 
\bigg[(E-Z) ^2  +
 i \ell_B ^{-2} ({\rm  Im\,}\,\bm{n}) \cdot \bm{\sigma }\bigg]\Psi.
\end{eqnarray*}
This implies
\begin{eqnarray*} 
\bigg[
(\bm{\pi}  ^\dagger c_0^2\Xi_0 \bm{\pi}) \, \sigma _0
-\ell_B^{-2}  \bm{n}\cdot \bm{\sigma } 
\bigg]\Psi
&=& 
(E-Z)^2\Psi.
\end{eqnarray*} 
Similarly to the case without magnetic field, 
one has
\begin{eqnarray*}
\big[c^2 \bm{\pi}^\dagger \Xi \bm{\pi} +2(E/\hbar) (X_0,Y_0)\bm{\pi}
\big] \sigma^0\Psi
- \ell_B^{-2} \bm{n}\cdot \bm{\sigma } \Psi &=& E^2\Psi,
\end{eqnarray*} 
which coincides with Eq.(\ref{eq:comp2}) in the text.

\section{Landau levels for an anisotropic mass}
\label{sec:aniso}
Let us summarize the standard Landau quantization of
electrons with parabolic dispersion with 
anisotropic masses (effective mass approximation) 
described by the following Hamiltonian
$$
H = 
 \mb{\pi} ^\dagger  \frac {1}{2m^*}  {\Xi}_L   \mb{\pi}, 
$$
with $\mb{\pi} = \mb{p}-e\mb{A} = \mb{\pi}^\dagger$, 
${\rm rot}\,  \mb{A}  = B \hat z$, and 
$$
\Xi_L   = 
\mmat
{\xi_x}
{\xi_{xy}}
{\xi_{xy}}
{\xi_y},
$$
where
\begin{alignat*}{1} 
\left(\frac {\ell_B}{\hbar}\right)^2[\pi_x, \pi_y] = i,
\ \
\ell_B = \sqrt{\frac {\hbar}{eB} }.
\end{alignat*} 
Here we have assumed $eB>0$ without loss of generality. 
Since the matrix $\Xi_L $ is real symmetric, it is diagonalized by
the orthogonal matrix $V$ as
\begin{alignat*}{1} 
\Xi_L  &= V ^\dagger   \Xi_D V, 
\\
\Xi_D &= \text{diag}(\xi_X,\xi_Y),\ \xi_X\xi_Y=\det \Xi_L ,\
 \xi_X+\xi_Y= {\rm Tr}\, \Xi_L,
\\
V &= 
\mmat
{\cos\theta}{-\sin\theta}
{\sin\theta}{\cos\theta}.\ ^\exists\theta\in \mathbb{R}
\end{alignat*} 
Then  we have
\begin{alignat*}{1} 
H = \mb{\Pi}  ^\dagger \Xi _D \mb{\Pi} ,\\
\mb{\Pi}  \equiv V \mb{\pi}, 
\\
\left(\frac {\ell_B}{\hbar}\right)^2[\Pi_X, \Pi_Y] =i.
\end{alignat*} 

Now defining a bosonic operator (with $[a,a ^\dagger ] = 1$),
\mynote{
\begin{alignat*}{1} 
a &= \frac {1}{\sqrt{2}} \frac {\ell_B}{\hbar} (\Pi_X+ i \Pi_Y)
\\
\Pi_X \frac {\ell_B}{\hbar} &= \frac {1}{\sqrt{2}} ( a+a ^\dagger )
\\
\Pi_Y \frac {\ell_B}{\hbar} &= \frac {1}{i\sqrt{2}} ( a-a ^\dagger )
\end{alignat*} 
\begin{alignat*}{1} 
\frac {1}{2i} [a+a ^\dagger , a- a ^\dagger ] &=  \frac {-1}{i}[ a, a ^\dagger ]=i
\\
[a,a ^\dagger ] &= 1
\end{alignat*} 
}
\begin{alignat*}{1} 
a &= \frac {1}{\sqrt{2}} \frac {\ell_B}{\hbar} (\Pi_X+ i \Pi_Y),
\end{alignat*} 
the Hamiltonian is written as
\begin{alignat*}{1} 
H &= 
\frac {\hbar\omega}{4}
\bigg[
\xi_X (a+a ^\dagger )^2
-
\xi_Y (a-a ^\dagger )^2\bigg]
\end{alignat*} 
with $\omega = eB/m^*$.

Now we define a new bosonic operator ($[b,b ^\dagger ]=1$)
as 
\begin{alignat*}{1} 
a &= u b + v^* b ^\dagger, 
\\
a ^\dagger  &= u^* b ^\dagger  + v b
\end{alignat*} 
with
$
[a, a ^\dagger  ] = 
[
 u b + v^* b ^\dagger ,
 u^* b ^\dagger  + v b ] = |u|^2-|v|^2=1.$  
Here we choose
\begin{eqnarray*} 
\xi_X(u+v)^2 &=&  \xi_Y (u-v)^2,
\\
u+v &=& C \sqrt{ \xi_Y},
\\
u-v &=&  -   C \sqrt{ \xi_X}.
\end{eqnarray*} 
Assuming $\xi_X,\xi_Y>0$ and imposing 
$|u|^2-|v|^2=1$, we have $|C|^2=1/\sqrt{\xi_X\xi_Y} =
1/(\det \mb{\Xi}_L )^{1/2}$ and therefore 
arrive at 
$$
u 
=\frac {\sqrt{\xi_X}+ \sqrt{\xi_Y}}{2(\det \mb{\Xi}_L )^{1/4}}, 
\quad 
v 
=\frac {-\sqrt{\xi_X}+ \sqrt{\xi_Y}}{2(\det \mb{\Xi}_L )^{1/4}}. 
$$

Finally, the  Hamiltonian is written as
\begin{alignat*}{1} 
H &= \frac {1}{2} {\hbar \omega} 
(b b ^\dagger +b ^\dagger b )
|C|^2 (\xi_X\xi_Y) = \hbar \omega _\Xi \left( b ^\dagger b + \frac {1}{2} \right),
\\
\omega_\Xi&=  {\omega }{\sqrt{\det \mb{\Xi}_L }} 
=\frac {eB}{m^*} {\sqrt{\det \mb{\Xi}_L }} 
=\frac {eB}{m^*} {\sqrt{\xi_X\xi_Y}}.
\end{alignat*} 

\section{Derivation of Eq.(\ref{chiral_basis})}
\label{equivalence}
The equation above Eq.(\ref{chiral_basis})
can be expressed, by 
introducing a dynamical momentum $\bm{\pi}_E' = \bm{\pi} +\Delta \bm{p}_E'$ in terms of 
a real vector $\Delta \bm{p}_E'$ satisfying 
a relation 
$\bm{\alpha}\cdot \Delta \bm{p}_E' = -E\beta$, as 
$$
 \left(\begin{array}{cc}
 mcc_r & \bm{\alpha}\cdot \bm{\pi}_E'\\
 \bm{\alpha}^*\cdot \bm{\pi}_E' & - m cc_r
 \end{array}\right)
 \left(\begin{array}{c}
 \psi_+^m \\
 \psi_-^m
 \end{array}\right) = E
 \left(\begin{array}{c}
 \psi_+^m \\
 \psi_-^m
 \end{array}\right).
$$
We can show that $ \Delta \bm{p}_E' =  \Delta \bm{p}_E$ 
by multiplying the matrix on the 
left-hand side  of the equation once again to get 
$$
[ {\rm Im}(\alpha_X\alpha_Y^*)(\bm{\pi}_E'^\dagger \Xi'  \bm{\pi}_E' \mp \hbar^2/\ell_B^2) +(mcc_r)^2] \psi_\pm^m=
  E^2\psi_\pm^m
$$
with 
$$
 \Xi'=\frac{1}{{\rm Im}(\alpha_X\alpha_Y^*)}\mmat{|\alpha_X|^2}{{\rm Re}(\alpha_X\alpha_Y^*)}{{\rm Re}(\alpha_X\alpha_Y^*)}{|\alpha_Y|^2}.
$$
Comparing this with Eq. (\ref{eq-sq}), we can see that $\Xi = \Xi'$, $\Delta \bm{p}_E = \Delta \bm{p}_E'$, and 
$ c_r^2={\rm Im}(\alpha_X\alpha_Y^*) $.




\begin{thebibliography}{50}%
\makeatletter
\providecommand \@ifxundefined [1]{%
 \@ifx{#1\undefined}
}%
\providecommand \@ifnum [1]{%
 \ifnum #1\expandafter \@firstoftwo
 \else \expandafter \@secondoftwo
 \fi
}%
\providecommand \@ifx [1]{%
 \ifx #1\expandafter \@firstoftwo
 \else \expandafter \@secondoftwo
 \fi
}%
\providecommand \natexlab [1]{#1}%
\providecommand \enquote  [1]{``#1''}%
\providecommand \bibnamefont  [1]{#1}%
\providecommand \bibfnamefont [1]{#1}%
\providecommand \citenamefont [1]{#1}%
\providecommand \href@noop [0]{\@secondoftwo}%
\providecommand \href [0]{\begingroup \@sanitize@url \@href}%
\providecommand \@href[1]{\@@startlink{#1}\@@href}%
\providecommand \@@href[1]{\endgroup#1\@@endlink}%
\providecommand \@sanitize@url [0]{\catcode `\\12\catcode `\$12\catcode
  `\&12\catcode `\#12\catcode `\^12\catcode `\_12\catcode `\%12\relax}%
\providecommand \@@startlink[1]{}%
\providecommand \@@endlink[0]{}%
\providecommand \url  [0]{\begingroup\@sanitize@url \@url }%
\providecommand \@url [1]{\endgroup\@href {#1}{\urlprefix }}%
\providecommand \urlprefix  [0]{URL }%
\providecommand \Eprint [0]{\href }%
\providecommand \doibase [0]{http://dx.doi.org/}%
\providecommand \selectlanguage [0]{\@gobble}%
\providecommand \bibinfo  [0]{\@secondoftwo}%
\providecommand \bibfield  [0]{\@secondoftwo}%
\providecommand \translation [1]{[#1]}%
\providecommand \BibitemOpen [0]{}%
\providecommand \bibitemStop [0]{}%
\providecommand \bibitemNoStop [0]{.\EOS\space}%
\providecommand \EOS [0]{\spacefactor3000\relax}%
\providecommand \BibitemShut  [1]{\csname bibitem#1\endcsname}%
\let\auto@bib@innerbib\@empty
\bibitem [{\citenamefont {Novoselov}\ \emph {et~al.}(2005)\citenamefont
  {Novoselov}, \citenamefont {Geim}, \citenamefont {Morozov}, \citenamefont
  {Jiang}, \citenamefont {Katsnelson}, \citenamefont {Grigorieva},
  \citenamefont {Dubonos},\ and\ \citenamefont {Firsov}}]{Novo05}%
  \BibitemOpen
  \bibfield  {author} {\bibinfo {author} {\bibfnamefont {K.~S.}\ \bibnamefont
  {Novoselov}}, \bibinfo {author} {\bibfnamefont {A.~K.}\ \bibnamefont {Geim}},
  \bibinfo {author} {\bibfnamefont {S.~V.}\ \bibnamefont {Morozov}}, \bibinfo
  {author} {\bibfnamefont {D.}~\bibnamefont {Jiang}}, \bibinfo {author}
  {\bibfnamefont {M.~I.}\ \bibnamefont {Katsnelson}}, \bibinfo {author}
  {\bibfnamefont {I.~V.}\ \bibnamefont {Grigorieva}}, \bibinfo {author}
  {\bibfnamefont {S.~V.}\ \bibnamefont {Dubonos}}, \ and\ \bibinfo {author}
  {\bibfnamefont {A.~A.}\ \bibnamefont {Firsov}},\ }\href
  {http://dx.doi.org/10.1038/nature04233} {\bibfield  {journal} {\bibinfo
  {journal} {Nature}\ }\textbf {\bibinfo {volume} {438}},\ \bibinfo {pages}
  {197} (\bibinfo {year} {2005})}\BibitemShut {NoStop}%
\bibitem [{\citenamefont {Aoki}\ and\ \citenamefont
  {Dresselhaus}(2014)}]{Aoki14}%
  \BibitemOpen
  \bibinfo {editor} {\bibfnamefont {H.}~\bibnamefont {Aoki}}\ and\ \bibinfo
  {editor} {\bibfnamefont {M.~S.}\ \bibnamefont {Dresselhaus}},\ eds.,\
  \href@noop {} {\emph {\bibinfo {title} {\it Physics of Graphene}}}\ (\bibinfo
   {publisher} {Springer},\ \bibinfo {year} {2014})\BibitemShut {NoStop}%
\bibitem [{\citenamefont {Harrison}(1989)}]{Harrison89}%
  \BibitemOpen
  \bibfield  {author} {\bibinfo {author} {\bibfnamefont {W.~A.}\ \bibnamefont
  {Harrison}},\ }\href@noop {} {\emph {\bibinfo {title} {\it Electronic
  Structure and Properties of Solids}}}\ (\bibinfo  {publisher} {Dover},\
  \bibinfo {year} {1989})\BibitemShut {NoStop}%
\bibitem [{\citenamefont {Kane}\ and\ \citenamefont {Mele}(2005)}]{Kane05}%
  \BibitemOpen
  \bibfield  {author} {\bibinfo {author} {\bibfnamefont {C.~L.}\ \bibnamefont
  {Kane}}\ and\ \bibinfo {author} {\bibfnamefont {E.~J.}\ \bibnamefont
  {Mele}},\ }\href {\doibase 10.1103/PhysRevLett.95.146802} {\bibfield
  {journal} {\bibinfo  {journal} {Phys. Rev. Lett.}\ }\textbf {\bibinfo
  {volume} {95}},\ \bibinfo {pages} {146802} (\bibinfo {year}
  {2005})}\BibitemShut {NoStop}%
\bibitem [{\citenamefont {Moore}(2010)}]{Moore10}%
  \BibitemOpen
  \bibfield  {author} {\bibinfo {author} {\bibfnamefont {J.~E.}\ \bibnamefont
  {Moore}},\ }\href {http://dx.doi.org/10.1038/nature08916} {\bibfield
  {journal} {\bibinfo  {journal} {Nature}\ }\textbf {\bibinfo {volume} {464}},\
  \bibinfo {pages} {194} (\bibinfo {year} {2010})}\BibitemShut {NoStop}%
\bibitem [{\citenamefont {Hasan}\ and\ \citenamefont {Kane}(2010)}]{Hasan10}%
  \BibitemOpen
  \bibfield  {author} {\bibinfo {author} {\bibfnamefont {M.~Z.}\ \bibnamefont
  {Hasan}}\ and\ \bibinfo {author} {\bibfnamefont {C.~L.}\ \bibnamefont
  {Kane}},\ }\href {\doibase 10.1103/RevModPhys.82.3045} {\bibfield  {journal}
  {\bibinfo  {journal} {Rev. Mod. Phys.}\ }\textbf {\bibinfo {volume} {82}},\
  \bibinfo {pages} {3045} (\bibinfo {year} {2010})}\BibitemShut {NoStop}%
\bibitem [{\citenamefont {Qi}\ and\ \citenamefont {Zhang}(2011)}]{RMPZhang11}%
  \BibitemOpen
  \bibfield  {author} {\bibinfo {author} {\bibfnamefont {X.-L.}\ \bibnamefont
  {Qi}}\ and\ \bibinfo {author} {\bibfnamefont {S.-C.}\ \bibnamefont {Zhang}},\
  }\href {\doibase 10.1103/RevModPhys.83.1057} {\bibfield  {journal} {\bibinfo
  {journal} {Rev. Mod. Phys.}\ }\textbf {\bibinfo {volume} {83}},\ \bibinfo
  {pages} {1057} (\bibinfo {year} {2011})}\BibitemShut {NoStop}%
\bibitem [{\citenamefont {Bernevig}\ \emph {et~al.}(2006)\citenamefont
  {Bernevig}, \citenamefont {Hughes},\ and\ \citenamefont
  {Zhang}}]{Bernevig06}%
  \BibitemOpen
  \bibfield  {author} {\bibinfo {author} {\bibfnamefont {B.~A.}\ \bibnamefont
  {Bernevig}}, \bibinfo {author} {\bibfnamefont {T.~L.}\ \bibnamefont
  {Hughes}}, \ and\ \bibinfo {author} {\bibfnamefont {S.~C.}\ \bibnamefont
  {Zhang}},\ }\href@noop {} {\bibfield  {journal} {\bibinfo  {journal}
  {Science}\ }\textbf {\bibinfo {volume} {314}},\ \bibinfo {pages} {1757}
  (\bibinfo {year} {2006})}\BibitemShut {NoStop}%
\bibitem [{\citenamefont {Hatsugai}\ \emph {et~al.}(1996)\citenamefont
  {Hatsugai}, \citenamefont {Kohmoto},\ and\ \citenamefont {Wu}}]{Hats96}%
  \BibitemOpen
  \bibfield  {author} {\bibinfo {author} {\bibfnamefont {Y.}~\bibnamefont
  {Hatsugai}}, \bibinfo {author} {\bibfnamefont {M.}~\bibnamefont {Kohmoto}}, \
  and\ \bibinfo {author} {\bibfnamefont {Y.-S.}\ \bibnamefont {Wu}},\ }\href
  {\doibase 10.1103/PhysRevB.54.4898} {\bibfield  {journal} {\bibinfo
  {journal} {Phys. Rev. B}\ }\textbf {\bibinfo {volume} {54}},\ \bibinfo
  {pages} {4898} (\bibinfo {year} {1996})}\BibitemShut {NoStop}%
\bibitem [{\citenamefont {Katayama}\ \emph {et~al.}(2006)\citenamefont
  {Katayama}, \citenamefont {Kobayashi},\ and\ \citenamefont
  {Suzumura}}]{Kata06}%
  \BibitemOpen
  \bibfield  {author} {\bibinfo {author} {\bibfnamefont {S.}~\bibnamefont
  {Katayama}}, \bibinfo {author} {\bibfnamefont {A.}~\bibnamefont {Kobayashi}},
  \ and\ \bibinfo {author} {\bibfnamefont {Y.}~\bibnamefont {Suzumura}},\
  }\href@noop {} {\bibfield  {journal} {\bibinfo  {journal} {J. Phys. Soc.
  Jpn.}\ }\textbf {\bibinfo {volume} {75}},\ \bibinfo {pages} {054705}
  (\bibinfo {year} {2006})}\BibitemShut {NoStop}%
\bibitem [{\citenamefont {Kobayashi}\ \emph {et~al.}(2007)\citenamefont
  {Kobayashi}, \citenamefont {Katayama}, \citenamefont {Suzumura},\ and\
  \citenamefont {Fukuyama}}]{Koba07}%
  \BibitemOpen
  \bibfield  {author} {\bibinfo {author} {\bibfnamefont {A.}~\bibnamefont
  {Kobayashi}}, \bibinfo {author} {\bibfnamefont {S.}~\bibnamefont {Katayama}},
  \bibinfo {author} {\bibfnamefont {Y.}~\bibnamefont {Suzumura}}, \ and\
  \bibinfo {author} {\bibfnamefont {H.}~\bibnamefont {Fukuyama}},\ }\href@noop
  {} {\bibfield  {journal} {\bibinfo  {journal} {J. Phys. Soc. Jpn.}\ }\textbf
  {\bibinfo {volume} {76}},\ \bibinfo {pages} {034711} (\bibinfo {year}
  {2007})}\BibitemShut {NoStop}%
\bibitem [{\citenamefont {Kajita}\ \emph {et~al.}(2014)\citenamefont {Kajita},
  \citenamefont {Nishio}, \citenamefont {Tajima}, \citenamefont {Suzumura},\
  and\ \citenamefont {Kobayashi}}]{kajita14}%
  \BibitemOpen
  \bibfield  {author} {\bibinfo {author} {\bibfnamefont {K.}~\bibnamefont
  {Kajita}}, \bibinfo {author} {\bibfnamefont {Y.}~\bibnamefont {Nishio}},
  \bibinfo {author} {\bibfnamefont {N.}~\bibnamefont {Tajima}}, \bibinfo
  {author} {\bibfnamefont {Y.}~\bibnamefont {Suzumura}}, \ and\ \bibinfo
  {author} {\bibfnamefont {A.}~\bibnamefont {Kobayashi}},\ }\href {\doibase
  10.7566/JPSJ.83.072002} {\bibfield  {journal} {\bibinfo  {journal} {Journal
  of the Physical Society of Japan}\ }\textbf {\bibinfo {volume} {83}},\
  \bibinfo {pages} {072002} (\bibinfo {year} {2014})}\BibitemShut {NoStop}%
\bibitem [{\citenamefont {Lee}(1993)}]{Lee93}%
  \BibitemOpen
  \bibfield  {author} {\bibinfo {author} {\bibfnamefont {P.~A.}\ \bibnamefont
  {Lee}},\ }\href {\doibase 10.1103/PhysRevLett.71.1887} {\bibfield  {journal}
  {\bibinfo  {journal} {Phys. Rev. Lett.}\ }\textbf {\bibinfo {volume} {71}},\
  \bibinfo {pages} {1887} (\bibinfo {year} {1993})}\BibitemShut {NoStop}%
\bibitem [{\citenamefont {Hatsugai}\ and\ \citenamefont
  {Lee}(1993)}]{PhysRevB.48.4204}%
  \BibitemOpen
  \bibfield  {author} {\bibinfo {author} {\bibfnamefont {Y.}~\bibnamefont
  {Hatsugai}}\ and\ \bibinfo {author} {\bibfnamefont {P.~A.}\ \bibnamefont
  {Lee}},\ }\href@noop {} {\bibfield  {journal} {\bibinfo  {journal} {Phys.
  Rev. B}\ }\textbf {\bibinfo {volume} {48}},\ \bibinfo {pages} {4204}
  (\bibinfo {year} {1993})}\BibitemShut {NoStop}%
\bibitem [{\citenamefont {Berry}(1984)}]{Berry84}%
  \BibitemOpen
  \bibfield  {author} {\bibinfo {author} {\bibfnamefont {M.~V.}\ \bibnamefont
  {Berry}},\ }\href@noop {} {\bibfield  {journal} {\bibinfo  {journal} {Proc.\
  R.\ Soc.}\ }\textbf {\bibinfo {volume} {A392}},\ \bibinfo {pages} {45}
  (\bibinfo {year} {1984})}\BibitemShut {NoStop}%
\bibitem [{\citenamefont {Herring}(1937)}]{Herring37}%
  \BibitemOpen
  \bibfield  {author} {\bibinfo {author} {\bibfnamefont {C.}~\bibnamefont
  {Herring}},\ }\href {\doibase 10.1103/PhysRev.52.365} {\bibfield  {journal}
  {\bibinfo  {journal} {Phys. Rev.}\ }\textbf {\bibinfo {volume} {52}},\
  \bibinfo {pages} {365} (\bibinfo {year} {1937})}\BibitemShut {NoStop}%
\bibitem [{\citenamefont {Hatsugai}(2010)}]{Hatsugai10S}%
  \BibitemOpen
  \bibfield  {author} {\bibinfo {author} {\bibfnamefont {Y.}~\bibnamefont
  {Hatsugai}},\ }\href@noop {} {\bibfield  {journal} {\bibinfo  {journal} {New
  J. Phys.}\ }\textbf {\bibinfo {volume} {12}},\ \bibinfo {pages} {065004}
  (\bibinfo {year} {2010})}\BibitemShut {NoStop}%
\bibitem [{\citenamefont {Hatsugai}\ and\ \citenamefont
  {Aoki}(2014)}]{Hatsbook14}%
  \BibitemOpen
  \bibfield  {author} {\bibinfo {author} {\bibfnamefont {Y.}~\bibnamefont
  {Hatsugai}}\ and\ \bibinfo {author} {\bibfnamefont {H.}~\bibnamefont
  {Aoki}},\ }\href@noop {} {\emph {\bibinfo {title} {\it Physics of
  Graphene}}},\ edited by\ \bibinfo {editor} {\bibfnamefont {H.}~\bibnamefont
  {Aoki}}\ and\ \bibinfo {editor} {\bibfnamefont {M.~S.}\ \bibnamefont
  {Dresselhaus}}\ (\bibinfo  {publisher} {Springer},\ \bibinfo {year} {2014})\
  p.\ \bibinfo {pages} {213}\BibitemShut {NoStop}%
\bibitem [{\citenamefont {Nielsen}\ and\ \citenamefont {Ninomiya}(1981)}]{NNT}%
  \BibitemOpen
  \bibfield  {author} {\bibinfo {author} {\bibfnamefont {H.~B.}\ \bibnamefont
  {Nielsen}}\ and\ \bibinfo {author} {\bibfnamefont {M.}~\bibnamefont
  {Ninomiya}},\ }\href@noop {} {\bibfield  {journal} {\bibinfo  {journal}
  {Nucl. Phys. B}\ }\textbf {\bibinfo {volume} {185}},\ \bibinfo {pages} {20}
  (\bibinfo {year} {1981})}\BibitemShut {NoStop}%
\bibitem [{\citenamefont {Hatsugai}\ \emph {et~al.}(2007)\citenamefont
  {Hatsugai}, \citenamefont {Fukui},\ and\ \citenamefont
  {Aoki}}]{HFAstability}%
  \BibitemOpen
  \bibfield  {author} {\bibinfo {author} {\bibfnamefont {Y.}~\bibnamefont
  {Hatsugai}}, \bibinfo {author} {\bibfnamefont {T.}~\bibnamefont {Fukui}}, \
  and\ \bibinfo {author} {\bibfnamefont {H.}~\bibnamefont {Aoki}},\ }\href@noop
  {} {\bibfield  {journal} {\bibinfo  {journal} {Euro. Phys. J. Spec. Top.}\
  }\textbf {\bibinfo {volume} {148}},\ \bibinfo {pages} {133} (\bibinfo {year}
  {2007})}\BibitemShut {NoStop}%
\bibitem [{\citenamefont {Hatsugai}(2009)}]{HatsuSSC}%
  \BibitemOpen
  \bibfield  {author} {\bibinfo {author} {\bibfnamefont {Y.}~\bibnamefont
  {Hatsugai}},\ }\href@noop {} {\bibfield  {journal} {\bibinfo  {journal}
  {Solid State Comm.}\ }\textbf {\bibinfo {volume} {149}},\ \bibinfo {pages}
  {1016} (\bibinfo {year} {2009})}\BibitemShut {NoStop}%
\bibitem [{\citenamefont {Witten}(1982)}]{Witten82}%
  \BibitemOpen
  \bibfield  {author} {\bibinfo {author} {\bibfnamefont {E.}~\bibnamefont
  {Witten}},\ }\href@noop {} {\bibfield  {journal} {\bibinfo  {journal} {Nucl.
  Phys. B}\ }\textbf {\bibinfo {volume} {202}},\ \bibinfo {pages} {253}
  (\bibinfo {year} {1982})}\BibitemShut {NoStop}%
\bibitem [{\citenamefont {Ezawa}(2007)}]{Ezawa07}%
  \BibitemOpen
  \bibfield  {author} {\bibinfo {author} {\bibfnamefont {M.}~\bibnamefont
  {Ezawa}},\ }\href@noop {} {\bibfield  {journal} {\bibinfo  {journal} {Physica
  E}\ }\textbf {\bibinfo {volume} {40}},\ \bibinfo {pages} {269} (\bibinfo
  {year} {2007})}\BibitemShut {NoStop}%
\bibitem [{\citenamefont {Kailasvuori}(2009)}]{EPLSUSY09}%
  \BibitemOpen
  \bibfield  {author} {\bibinfo {author} {\bibfnamefont {J.}~\bibnamefont
  {Kailasvuori}},\ }\href@noop {} {\bibfield  {journal} {\bibinfo  {journal}
  {Euro. Phys. Lett.}\ }\textbf {\bibinfo {volume} {87}},\ \bibinfo {pages}
  {47008} (\bibinfo {year} {2009})}\BibitemShut {NoStop}%
\bibitem [{\citenamefont {Hatsugai}\ \emph {et~al.}(2006)\citenamefont
  {Hatsugai}, \citenamefont {Fukui},\ and\ \citenamefont
  {Aoki}}]{Hatsugai06-Gra}%
  \BibitemOpen
  \bibfield  {author} {\bibinfo {author} {\bibfnamefont {Y.}~\bibnamefont
  {Hatsugai}}, \bibinfo {author} {\bibfnamefont {T.}~\bibnamefont {Fukui}}, \
  and\ \bibinfo {author} {\bibfnamefont {H.}~\bibnamefont {Aoki}},\ }\href
  {\doibase 10.1103/PhysRevB.74.205414} {\bibfield  {journal} {\bibinfo
  {journal} {Phys. Rev. B}\ }\textbf {\bibinfo {volume} {74}},\ \bibinfo
  {pages} {205414} (\bibinfo {year} {2006})}\BibitemShut {NoStop}%
\bibitem [{\citenamefont {Kawarabayashi}\ \emph {et~al.}(2009)\citenamefont
  {Kawarabayashi}, \citenamefont {Hatsugai},\ and\ \citenamefont
  {Aoki}}]{Kawarabayashi09}%
  \BibitemOpen
  \bibfield  {author} {\bibinfo {author} {\bibfnamefont {T.}~\bibnamefont
  {Kawarabayashi}}, \bibinfo {author} {\bibfnamefont {Y.}~\bibnamefont
  {Hatsugai}}, \ and\ \bibinfo {author} {\bibfnamefont {H.}~\bibnamefont
  {Aoki}},\ }\href {\doibase 10.1103/PhysRevLett.103.156804} {\bibfield
  {journal} {\bibinfo  {journal} {Phys. Rev. Lett.}\ }\textbf {\bibinfo
  {volume} {103}},\ \bibinfo {pages} {156804} (\bibinfo {year}
  {2009})}\BibitemShut {NoStop}%
\bibitem [{\citenamefont {Ryu}\ and\ \citenamefont {Hatsugai}(2002)}]{Ryu02}%
  \BibitemOpen
  \bibfield  {author} {\bibinfo {author} {\bibfnamefont {S.}~\bibnamefont
  {Ryu}}\ and\ \bibinfo {author} {\bibfnamefont {Y.}~\bibnamefont {Hatsugai}},\
  }\href@noop {} {\bibfield  {journal} {\bibinfo  {journal} {Phys.\ Rev.\
  Lett.}\ }\textbf {\bibinfo {volume} {89}},\ \bibinfo {pages} {077002}
  (\bibinfo {year} {2002})}\BibitemShut {NoStop}%
\bibitem [{\citenamefont {Kawarabayashi}\ \emph {et~al.}(2011)\citenamefont
  {Kawarabayashi}, \citenamefont {Hatsugai}, \citenamefont {Morimoto},\ and\
  \citenamefont {Aoki}}]{Kawa11}%
  \BibitemOpen
  \bibfield  {author} {\bibinfo {author} {\bibfnamefont {T.}~\bibnamefont
  {Kawarabayashi}}, \bibinfo {author} {\bibfnamefont {Y.}~\bibnamefont
  {Hatsugai}}, \bibinfo {author} {\bibfnamefont {T.}~\bibnamefont {Morimoto}},
  \ and\ \bibinfo {author} {\bibfnamefont {H.}~\bibnamefont {Aoki}},\ }\href
  {\doibase 10.1103/PhysRevB.83.153414} {\bibfield  {journal} {\bibinfo
  {journal} {Phys. Rev. B}\ }\textbf {\bibinfo {volume} {83}},\ \bibinfo
  {pages} {153414} (\bibinfo {year} {2011})}\BibitemShut {NoStop}%
\bibitem [{\citenamefont {Suzuki}\ \emph {et~al.}(2014)\citenamefont {Suzuki},
  \citenamefont {Sakano}, \citenamefont {Zhang}, \citenamefont {Akashi},
  \citenamefont {Morikawa}, \citenamefont {Harasawa}, \citenamefont {Yaji},
  \citenamefont {Kuroda}, \citenamefont {Miyamoto}, \citenamefont {Okuda},
  \citenamefont {Ishizaka}, \citenamefont {Arita},\ and\ \citenamefont
  {Iwasa}}]{mos}%
  \BibitemOpen
  \bibfield  {author} {\bibinfo {author} {\bibfnamefont {R.}~\bibnamefont
  {Suzuki}}, \bibinfo {author} {\bibfnamefont {M.}~\bibnamefont {Sakano}},
  \bibinfo {author} {\bibfnamefont {Y.~J.}\ \bibnamefont {Zhang}}, \bibinfo
  {author} {\bibfnamefont {R.}~\bibnamefont {Akashi}}, \bibinfo {author}
  {\bibfnamefont {D.}~\bibnamefont {Morikawa}}, \bibinfo {author}
  {\bibfnamefont {A.}~\bibnamefont {Harasawa}}, \bibinfo {author}
  {\bibfnamefont {K.}~\bibnamefont {Yaji}}, \bibinfo {author} {\bibfnamefont
  {K.}~\bibnamefont {Kuroda}}, \bibinfo {author} {\bibfnamefont
  {K.}~\bibnamefont {Miyamoto}}, \bibinfo {author} {\bibfnamefont
  {T.}~\bibnamefont {Okuda}}, \bibinfo {author} {\bibfnamefont
  {K.}~\bibnamefont {Ishizaka}}, \bibinfo {author} {\bibfnamefont
  {R.}~\bibnamefont {Arita}}, \ and\ \bibinfo {author} {\bibfnamefont
  {Y.}~\bibnamefont {Iwasa}},\ }\href
  {http://dx.doi.org/10.1038/nnano.2014.148} {\bibfield  {journal} {\bibinfo
  {journal} {Nat Nano}\ }\textbf {\bibinfo {volume} {9}},\ \bibinfo {pages}
  {611} (\bibinfo {year} {2014})}\BibitemShut {NoStop}%
\bibitem [{\citenamefont {Kobayashi}\ \emph {et~al.}(2005)\citenamefont
  {Kobayashi}, \citenamefont {Katayama},\ and\ \citenamefont
  {Suzumura}}]{doi:10.1143/JPSJ.74.2897}%
  \BibitemOpen
  \bibfield  {author} {\bibinfo {author} {\bibfnamefont {A.}~\bibnamefont
  {Kobayashi}}, \bibinfo {author} {\bibfnamefont {S.}~\bibnamefont {Katayama}},
  \ and\ \bibinfo {author} {\bibfnamefont {Y.}~\bibnamefont {Suzumura}},\
  }\href {\doibase 10.1143/JPSJ.74.2897} {\bibfield  {journal} {\bibinfo
  {journal} {Journal of the Physical Society of Japan}\ }\textbf {\bibinfo
  {volume} {74}},\ \bibinfo {pages} {2897} (\bibinfo {year}
  {2005})}\BibitemShut {NoStop}%
\bibitem [{\citenamefont {Morinari}\ and\ \citenamefont
  {Suzumura}(2014)}]{doi:10.7566/JPSJ.83.094701}%
  \BibitemOpen
  \bibfield  {author} {\bibinfo {author} {\bibfnamefont {T.}~\bibnamefont
  {Morinari}}\ and\ \bibinfo {author} {\bibfnamefont {Y.}~\bibnamefont
  {Suzumura}},\ }\href@noop {} {\bibfield  {journal} {\bibinfo  {journal}
  {Journal of the Physical Society of Japan}\ }\textbf {\bibinfo {volume}
  {83}},\ \bibinfo {pages} {094701} (\bibinfo {year} {2014})}\BibitemShut
  {NoStop}%
\bibitem [{\citenamefont {Watanabe}\ \emph {et~al.}(2010)\citenamefont
  {Watanabe}, \citenamefont {Hatsugai},\ and\ \citenamefont
  {Aoki}}]{PhysRevB.82.241403}%
  \BibitemOpen
  \bibfield  {author} {\bibinfo {author} {\bibfnamefont {H.}~\bibnamefont
  {Watanabe}}, \bibinfo {author} {\bibfnamefont {Y.}~\bibnamefont {Hatsugai}},
  \ and\ \bibinfo {author} {\bibfnamefont {H.}~\bibnamefont {Aoki}},\ }\href
  {\doibase 10.1103/PhysRevB.82.241403} {\bibfield  {journal} {\bibinfo
  {journal} {Phys. Rev. B}\ }\textbf {\bibinfo {volume} {82}},\ \bibinfo
  {pages} {241403} (\bibinfo {year} {2010})}\BibitemShut {NoStop}%
\bibitem [{\citenamefont {Tarruell}\ \emph {et~al.}(2012)\citenamefont
  {Tarruell}, \citenamefont {Greif}, \citenamefont {Uehlinger}, \citenamefont
  {Jotzu},\ and\ \citenamefont {Esslinger}}]{movedDirac}%
  \BibitemOpen
  \bibfield  {author} {\bibinfo {author} {\bibfnamefont {L.}~\bibnamefont
  {Tarruell}}, \bibinfo {author} {\bibfnamefont {D.}~\bibnamefont {Greif}},
  \bibinfo {author} {\bibfnamefont {T.}~\bibnamefont {Uehlinger}}, \bibinfo
  {author} {\bibfnamefont {G.}~\bibnamefont {Jotzu}}, \ and\ \bibinfo {author}
  {\bibfnamefont {T.}~\bibnamefont {Esslinger}},\ }\href
  {http://dx.doi.org/10.1038/nature10871} {\bibfield  {journal} {\bibinfo
  {journal} {Nature}\ }\textbf {\bibinfo {volume} {483}},\ \bibinfo {pages}
  {302} (\bibinfo {year} {2012})}\BibitemShut {NoStop}%
\bibitem [{\citenamefont {Goldman}\ \emph {et~al.}(2009)\citenamefont
  {Goldman}, \citenamefont {Kubasiak}, \citenamefont {Bermudez}, \citenamefont
  {Gaspard}, \citenamefont {Lewenstein},\ and\ \citenamefont
  {Martin-Delgado}}]{PhysRevLett.103.035301}%
  \BibitemOpen
  \bibfield  {author} {\bibinfo {author} {\bibfnamefont {N.}~\bibnamefont
  {Goldman}}, \bibinfo {author} {\bibfnamefont {A.}~\bibnamefont {Kubasiak}},
  \bibinfo {author} {\bibfnamefont {A.}~\bibnamefont {Bermudez}}, \bibinfo
  {author} {\bibfnamefont {P.}~\bibnamefont {Gaspard}}, \bibinfo {author}
  {\bibfnamefont {M.}~\bibnamefont {Lewenstein}}, \ and\ \bibinfo {author}
  {\bibfnamefont {M.}~\bibnamefont {Martin-Delgado}},\ }\href {\doibase
  10.1103/PhysRevLett.103.035301} {\bibfield  {journal} {\bibinfo  {journal}
  {Phys. Rev. Lett.}\ }\textbf {\bibinfo {volume} {103}},\ \bibinfo {pages}
  {035301} (\bibinfo {year} {2009})}\BibitemShut {NoStop}%
\bibitem [{\citenamefont {Mei}\ \emph {et~al.}(2011)\citenamefont {Mei},
  \citenamefont {Zhu}, \citenamefont {Feng}, \citenamefont {Zhang},\ and\
  \citenamefont {Oh}}]{PhysRevA.84.023622}%
  \BibitemOpen
  \bibfield  {author} {\bibinfo {author} {\bibfnamefont {F.}~\bibnamefont
  {Mei}}, \bibinfo {author} {\bibfnamefont {S.-L.}\ \bibnamefont {Zhu}},
  \bibinfo {author} {\bibfnamefont {X.-L.}\ \bibnamefont {Feng}}, \bibinfo
  {author} {\bibfnamefont {Z.-M.}\ \bibnamefont {Zhang}}, \ and\ \bibinfo
  {author} {\bibfnamefont {C.}~\bibnamefont {Oh}},\ }\href {\doibase
  10.1103/PhysRevA.84.023622} {\bibfield  {journal} {\bibinfo  {journal} {Phys.
  Rev. A}\ }\textbf {\bibinfo {volume} {84}},\ \bibinfo {pages} {023622}
  (\bibinfo {year} {2011})}\BibitemShut {NoStop}%
\bibitem [{\citenamefont {Kawarabayashi}\ \emph {et~al.}(2013)\citenamefont
  {Kawarabayashi}, \citenamefont {Honda}, \citenamefont {Aoki},\ and\
  \citenamefont {Hatsugai}}]{Kawa13}%
  \BibitemOpen
  \bibfield  {author} {\bibinfo {author} {\bibfnamefont {T.}~\bibnamefont
  {Kawarabayashi}}, \bibinfo {author} {\bibfnamefont {T.}~\bibnamefont
  {Honda}}, \bibinfo {author} {\bibfnamefont {H.}~\bibnamefont {Aoki}}, \ and\
  \bibinfo {author} {\bibfnamefont {Y.}~\bibnamefont {Hatsugai}},\ }\href
  {\doibase 10.1063/1.4848396} {\bibfield  {journal} {\bibinfo  {journal} {AIP
  Conference Proceedings}\ }\textbf {\bibinfo {volume} {1566}},\ \bibinfo
  {pages} {283} (\bibinfo {year} {2013})}\BibitemShut {NoStop}%
\bibitem [{\citenamefont {Hatsugai}\ and\ \citenamefont
  {Kohmoto}(1990)}]{Hatsugai90}%
  \BibitemOpen
  \bibfield  {author} {\bibinfo {author} {\bibfnamefont {Y.}~\bibnamefont
  {Hatsugai}}\ and\ \bibinfo {author} {\bibfnamefont {M.}~\bibnamefont
  {Kohmoto}},\ }\href@noop {} {\bibfield  {journal} {\bibinfo  {journal} {Phys.
  \ Rev.\ B}\ }\textbf {\bibinfo {volume} {42}},\ \bibinfo {pages} {8282}
  (\bibinfo {year} {1990})}\BibitemShut {NoStop}%
\bibitem [{\citenamefont {Hatsugai}(2011)}]{hatsugai2011}%
  \BibitemOpen
  \bibfield  {author} {\bibinfo {author} {\bibfnamefont {Y.}~\bibnamefont
  {Hatsugai}},\ }\href@noop {} {\bibfield  {journal} {\bibinfo  {journal} {J.
  Phys. : Conf. Series}\ }\textbf {\bibinfo {volume} {334}},\ \bibinfo {pages}
  {012004} (\bibinfo {year} {2011})}\BibitemShut {NoStop}%
\bibitem [{\citenamefont {Kawarabayashi}\ \emph {et~al.}(2012)\citenamefont
  {Kawarabayashi}, \citenamefont {Hatsugai}, \citenamefont {Morimoto},\ and\
  \citenamefont {Aoki}}]{Kawa12}%
  \BibitemOpen
  \bibfield  {author} {\bibinfo {author} {\bibfnamefont {T.}~\bibnamefont
  {Kawarabayashi}}, \bibinfo {author} {\bibfnamefont {Y.}~\bibnamefont
  {Hatsugai}}, \bibinfo {author} {\bibfnamefont {T.}~\bibnamefont {Morimoto}},
  \ and\ \bibinfo {author} {\bibfnamefont {H.}~\bibnamefont {Aoki}},\ }\href
  {\doibase 10.1142/S2010194512006046} {\bibfield  {journal} {\bibinfo
  {journal} {Int. J. Mod. Phys.: Conf. Series}\ }\textbf {\bibinfo {volume}
  {11}},\ \bibinfo {pages} {145} (\bibinfo {year} {2012})}\BibitemShut
  {NoStop}%
\bibitem [{\citenamefont {Goerbig}\ \emph {et~al.}(2008)\citenamefont
  {Goerbig}, \citenamefont {Fuchs}, \citenamefont {Montambaux},\ and\
  \citenamefont {Pi\'echon}}]{Goerb08}%
  \BibitemOpen
  \bibfield  {author} {\bibinfo {author} {\bibfnamefont {M.~O.}\ \bibnamefont
  {Goerbig}}, \bibinfo {author} {\bibfnamefont {J.-N.}\ \bibnamefont {Fuchs}},
  \bibinfo {author} {\bibfnamefont {G.}~\bibnamefont {Montambaux}}, \ and\
  \bibinfo {author} {\bibfnamefont {F.}~\bibnamefont {Pi\'echon}},\ }\href
  {\doibase 10.1103/PhysRevB.78.045415} {\bibfield  {journal} {\bibinfo
  {journal} {Phys. Rev. B}\ }\textbf {\bibinfo {volume} {78}},\ \bibinfo
  {pages} {045415} (\bibinfo {year} {2008})}\BibitemShut {NoStop}%
\bibitem [{\citenamefont {Morinari}\ \emph {et~al.}(2009)\citenamefont
  {Morinari}, \citenamefont {Himura},\ and\ \citenamefont {Tohyama}}]{Mori09}%
  \BibitemOpen
  \bibfield  {author} {\bibinfo {author} {\bibfnamefont {T.}~\bibnamefont
  {Morinari}}, \bibinfo {author} {\bibfnamefont {T.}~\bibnamefont {Himura}}, \
  and\ \bibinfo {author} {\bibfnamefont {T.}~\bibnamefont {Tohyama}},\
  }\href@noop {} {\bibfield  {journal} {\bibinfo  {journal} {J. Phys. Soc.
  Jpn.}\ }\textbf {\bibinfo {volume} {78}},\ \bibinfo {pages} {023704}
  (\bibinfo {year} {2009})}\BibitemShut {NoStop}%
\bibitem [{\citenamefont {Morinari}\ and\ \citenamefont
  {Tohyama}(2010)}]{Mori10}%
  \BibitemOpen
  \bibfield  {author} {\bibinfo {author} {\bibfnamefont {T.}~\bibnamefont
  {Morinari}}\ and\ \bibinfo {author} {\bibfnamefont {T.}~\bibnamefont
  {Tohyama}},\ }\href@noop {} {\bibfield  {journal} {\bibinfo  {journal} {J.
  Phys. Soc. Jpn.}\ }\textbf {\bibinfo {volume} {79}},\ \bibinfo {pages}
  {044708} (\bibinfo {year} {2010})}\BibitemShut {NoStop}%
\bibitem [{\citenamefont {Aharonov}\ and\ \citenamefont {Casher}(1979)}]{ac79}%
  \BibitemOpen
  \bibfield  {author} {\bibinfo {author} {\bibfnamefont {Y.}~\bibnamefont
  {Aharonov}}\ and\ \bibinfo {author} {\bibfnamefont {A.}~\bibnamefont
  {Casher}},\ }\href {\doibase 10.1103/PhysRevA.19.2461} {\bibfield  {journal}
  {\bibinfo  {journal} {Phys. Rev. A}\ }\textbf {\bibinfo {volume} {19}},\
  \bibinfo {pages} {2461} (\bibinfo {year} {1979})}\BibitemShut {NoStop}%
\bibitem [{\citenamefont {Ludwig}\ \emph {et~al.}(1994)\citenamefont {Ludwig},
  \citenamefont {Fisher}, \citenamefont {Shankar},\ and\ \citenamefont
  {Grinstein}}]{Ludwig94}%
  \BibitemOpen
  \bibfield  {author} {\bibinfo {author} {\bibfnamefont {A.~W.~W.}\
  \bibnamefont {Ludwig}}, \bibinfo {author} {\bibfnamefont {M.~P.~A.}\
  \bibnamefont {Fisher}}, \bibinfo {author} {\bibfnamefont {R.}~\bibnamefont
  {Shankar}}, \ and\ \bibinfo {author} {\bibfnamefont {G.}~\bibnamefont
  {Grinstein}},\ }\href {\doibase 10.1103/PhysRevB.50.7526} {\bibfield
  {journal} {\bibinfo  {journal} {Phys. Rev. B}\ }\textbf {\bibinfo {volume}
  {50}},\ \bibinfo {pages} {7526} (\bibinfo {year} {1994})}\BibitemShut
  {NoStop}%
\bibitem [{\citenamefont {Haldane}(1988)}]{Haldane88}%
  \BibitemOpen
  \bibfield  {author} {\bibinfo {author} {\bibfnamefont {F.~D.~M.}\
  \bibnamefont {Haldane}},\ }\href {\doibase 10.1103/PhysRevLett.61.2015}
  {\bibfield  {journal} {\bibinfo  {journal} {Phys. Rev. Lett.}\ }\textbf
  {\bibinfo {volume} {61}},\ \bibinfo {pages} {2015} (\bibinfo {year}
  {1988})}\BibitemShut {NoStop}%
\bibitem [{\citenamefont {Fuchs}\ \emph {et~al.}(2010)\citenamefont {Fuchs},
  \citenamefont {Pi\'{e}chon}, \citenamefont {Goerbig},\ and\ \citenamefont
  {Montambaux}}]{Fuchs}%
  \BibitemOpen
  \bibfield  {author} {\bibinfo {author} {\bibfnamefont {J.~N.}\ \bibnamefont
  {Fuchs}}, \bibinfo {author} {\bibfnamefont {F.}~\bibnamefont {Pi\'{e}chon}},
  \bibinfo {author} {\bibfnamefont {M.~O.}\ \bibnamefont {Goerbig}}, \ and\
  \bibinfo {author} {\bibfnamefont {G.}~\bibnamefont {Montambaux}},\ }\href
  {\doibase 10.1140/epjb/e2010-00259-2} {\bibfield  {journal} {\bibinfo
  {journal} {The European Physical Journal B}\ }\textbf {\bibinfo {volume}
  {77}},\ \bibinfo {pages} {351} (\bibinfo {year} {2010})}\BibitemShut
  {NoStop}%
  \mcr{
  \bibitem [{\citenamefont {Semenoff}(1984)}]{PhysRevLett.53.2449}%
  \BibitemOpen
  \bibfield  {author} {\bibinfo {author} {\bibfnamefont {G.~W.}\ \bibnamefont
  {Semenoff}},\ }\href {\doibase 10.1103/PhysRevLett.53.2449} {\bibfield
  {journal} {\bibinfo  {journal} {Phys. Rev. Lett.}\ }\textbf {\bibinfo
  {volume} {53}},\ \bibinfo {pages} {2449} (\bibinfo {year}
  {1984})}\BibitemShut {NoStop}%
\bibitem [{\citenamefont {Jackiw}(1984)}]{PhysRevD.29.2375}%
  \BibitemOpen
  \bibfield  {author} {\bibinfo {author} {\bibfnamefont {R.}~\bibnamefont
  {Jackiw}},\ }\href {\doibase 10.1103/PhysRevD.29.2375} {\bibfield  {journal}
  {\bibinfo  {journal} {Phys. Rev. D}\ }\textbf {\bibinfo {volume} {29}},\
  \bibinfo {pages} {2375} (\bibinfo {year} {1984})}\BibitemShut {NoStop}%
  }
\bibitem [{Note1()}]{Note1}%
  \BibitemOpen
  \bibinfo {note} {Using a well known formula, we have $ (H_C^0)^2 = \hbar
  ^{-2} {[}\protect \bm {\sigma }\cdot (\protect \bm {X}, \protect \bm {Y}
  )\protect \bm {p} ] ^ 2 = \hbar ^{-2}{[} (\protect \bm {\sigma }\cdot
  \protect \bm {X}) p_x + (\protect \bm {\sigma }\cdot \protect \bm {Y}) p_y
  {\setbox \z@ \hbox {\frozen@everymath \@emptytoks \mathsurround \z@
  $\nulldelimiterspace \z@ \left ]\vcenter to\@ne \big@size {}\right .$}\box
  \z@ }^2 = \hbar ^{-2}(\protect \bm {X} \cdot \protect \bm {X} p_x^2 +
  2\protect \bm {X} \cdot \protect \bm {Y} p_xp_y + \protect \bm {Y} \cdot
  \protect \bm {Y} p_y^2) \sigma _0 = | (\protect \bm {X},\protect \bm {Y})
  \protect \bm {p} |^2 \sigma _0 = [ c_0^2 \protect \bm {p} ^\dagger \Xi _0
  \protect \bm {p}] \protect \tmspace +\thinmuskip {.1667em} \sigma _0
  $.}\BibitemShut {Stop}%
\bibitem [{Note2()}]{Note2}%
  \BibitemOpen
  \bibinfo {note} {$ \protect \qopname \relax m{det}{\Xi }_0 = \protect \frac
  {1}{(c_0\hbar )^4} {\setbox \z@ \hbox {\frozen@everymath \@emptytoks
  \mathsurround \z@ $\nulldelimiterspace \z@ \left (\vcenter to\@ne \big@size
  {}\right .$}\box \z@ }|\protect \bm {X} |^2|\protect \bm {Y} |^2-(\protect
  \bm {X}\cdot \protect \bm {Y} )^2{\setbox \z@ \hbox {\frozen@everymath
  \@emptytoks \mathsurround \z@ $\nulldelimiterspace \z@ \left )\vcenter to\@ne
  \big@size {}\right .$}\box \z@ }= \protect \frac {1}{(c_0\hbar )^4} |\protect
  \bm {X}\times \protect \bm {Y}|^2=1 $}\BibitemShut {NoStop}%
\end{thebibliography}

%


\end{document}